# Collaborative Optimization of Multi-microgrids System with Shared Energy Storage Based on Multi-agent Stochastic Game and Reinforcement Learning

Yijian Wang [1], Yang Cui [*,1], Yang Li [1], Yang Xu [1]

[1] *Key Laboratory of Modern Power System Simulation and Control & Renewable Energy Technology, Ministry of Education (Northeast Electric Power University), Jilin132012, China*

**Abstract**

Achieving the economical and stable operation of Multi-microgrids (MMG) systems is vital. However, there are still some challenging problems to be solved. Firstly, from the perspective of stable operation, it is necessary to minimize the energy fluctuation of the main grid. Secondly, the characteristics of energy conversion equipment need to be considered. Finally, privacy protection while reducing the operating cost of an MMG system is crucial. To address these challenges, a Data-driven strategy for MMG systems with Shared Energy Storage (SES) is proposed. The Mixed-Attention is applied to fit the conditions of the equipment, additionally, Multi-Agent Soft Actor-Critic(MA-SAC) and (Multi-Agent Win or Learn Fast Policy Hill-Climbing)MA-WoLF-PHC are proposed to solve the partially observable dynamic stochastic game problem. By testing the operation data of the MMG system in Northwest China, following conclusions are drawn: the R-Square ($R^2$) values of results reach 0.999, indicating the neural network effectively models the nonlinear conditions. The proposed MMG system framework can reduce energy fluctuations in the main grid by 1746.5kW in 24 hours and achieve a cost reduction of 16.21% in the test. Finally, the superiority of the proposed algorithms is verified through their fast convergence speed and excellent optimization performance.

## 1. Introductions

*1.1. Background and motivation*

Since the Second Industrial Revolution, electricity has become the dominant energy source for the development of human society. As a reliable primary energy source for electricity production, fossil fuels have a very high strategic position. However, because of the amount of carbon dioxide that's produced during their combustion, global climate change is caused, which leads to the greenhouse effect and have a huge impact on the ecological environment. To alleviate the predicament, renewable energy sources such as wind, solar, and biomass have received extensive attention. Due to the different distribution of renewable energy in different regions [1], to make full use of these renewable energy sources (RES) and reduce the operating cost of the entire power system [2-4], scholars have proposed the concept of the microgrid (MG). Therefore, the promotion of MG is crucial for achieving sustainable development. Up to now, many scholars have studied the related problems of MG operation, but there are still few studies on the MMG system. Especially when considering the privacy protection between MGs and the nonlinear situation of energy conversion equipment at the same time, conventional methods are often difficult to solve. This problem hinders the promotion and application of MG.

*1.2. Literature review*

MGs have two operation modes: islanded mode [5] and grid-connected operation mode [6]. The former can help solve power supply problems in remote areas and contributes to the speed of recovery after faults [7]. Under the grid-connected operation mode, the MG can exchange energy with other MGs and the distribution network, and the energy supply is more stable [8], more importantly, with the development of smart grids [9], the collaboration of multiple MGs will become the main operating mode in the future [10-11]. Therefore, the research on these two operating modes is necessary and meaningful but considering the wider application prospect of the former, this paper only takes the latter as the main research object.

| Nomenclature | | | |
|---|---|---|---|
| *Abbreviations* | | | |
| IEHS | Integrated electricity and heat system | $CR$ | Capacity Retention |
| CHP | Combined heat and power | $T$ | Cycling times of charge-discharge |
| BIGCC | Biomass integrated gasification combined cycle | $COST_t^{MGi}$ | Total cost of MG$i$ at time $t$ |
| RES | Renewable energy source | $COST_t^{SES}$ | Cost of Shared energy storage at time $t$ |
| GT | Gas Turbine | $R_{H-P}$ | Heat to power ration |
| ST | Steam Turbine | $S_t^{main,e}$ | Price of selling electricity to the main grid at time $t$ |
| SO | Stochastic optimization | $B_t^{main,e}$ | Price of buying electricity to the main grid at time $t$ |
| RO | Robust optimization | $S_t^{micro,e}$ | Price of selling electricity to microgrids at time $t$ |
| SOC | State of charge | $B_t^{micro,e}$ | Price of buying electricity to microgrids at time $t$ |
| MG | Microgrid | $S_t^{main,h}$ | Price of selling heat to the main grid at time $t$ |
| OEM | Optimal energy management | $B_t^{main,h}$ | Price of buying heat to the main grid at time $t$ |
| MMG | Multiple-Microgrids | $S_t^{micro,h}$ | Price of selling heat to microgrids at time $t$ |
| DP | Dynamic programming | $B_t^{micro,h}$ | Price of buying heat to microgrids at time $t$ |
| RL | Reinforcement learning | SOC | State of charge (%) |
| CCPP | Combined cycle power plant | $N_i$ | Number of Capstone C200 in MGi |
| DRL | Deep reinforcement learning | $P_t^{ij,e}$ | Electricity from MGi to MGj at time $t$ |
| SES | Shared energy storage | $H_t^{ij,h}$ | Heat from MGi to MGj at time $t$ |
| CTDE | Centralized training with decentralized execution | $Bid_i$ | Bid of MGi at time $t$ |
| MA | Muti-Agent. | $P_t^{mi,e}$ | Electricity from the main grid to MGi at time $t$ |
| ORC | Organic Rankine Cycle | $H_t^{mi,h}$ | Heat from the main grid to MGi at time $t$ |
| A-C | Actor-Critic | $w$ | Parameters of critic |
| DDPG | Deep deterministic policy gradient | $\theta$ | Parameters of actor |
| TD3 | Twin Delayed Deep Deterministic policy gradient algorithm | $\eta_{ST}$ | Efficiency of Steam Turbine |
| PPO | Proximal Policy Optimization | CCPP | Combined cycle power plant |
| SAC | Soft Actor-Critic | $\eta_{CCPP}$ | Efficiency of CCPP |
| HRSG | Heat Recovery Steam Generator | $\tau$ | Rate of the soft update |
| $\eta_{C200-N}$ | Efficiency of Capstone C200 | $f_\theta(\square)$ | The rule of reparameterization |
| $\eta_{HR}$ | Efficiency of HRSG | $\partial$ | Partial differentiation |
| $\eta_{BIGCC}$ | Efficiency of BIGCC | $\lambda$ | Learning rate |

In practical applications, OEM problems are usually transformed into mathematical optimization problems. Robust optimization (RO) can fully consider the uncertainty of the variables and describe them in a set, the solution results of RO are strictly constrained and have strong robustness, but because of this situation, the solution results are often very conservative [12-14]. When the distribution models of uncertain variables in microgrid dispatching problems are known, stochastic optimization (SO) can also be used to formulate a feasible dispatch scheme [15-16]. Distributionally robust optimization (DRO) can combine the reliability of robust optimization with the flexibility of random optimization, providing a more reasonable solution for the dispatching problem of MGs [17]. In addition, some swarm intelligence algorithms (such as Particle Swarm Optimization (PSO)[18]) are also used for OEM. Ref. [19] proposes to use of dynamic programming (DP) to solve the OEM of reversible solid oxide cell-based MGs.

With the application of MGs, MMG systems have received attention from the academic community as never before. In MMG, individual MGs can exchange energy efficiently with each other and the main grid, this not only helps to solve the burden on the main grid but also improves the reliability of MG. According to this, Ref.[20] is considered that the model-based method suitable for solving MMG energy management is mainly: Dual decomposition, Analytical Target Cascading(ATC), Alternation Direction Method of Multipliers(ADMM), and Nash equilibrium In Ref.[21], a novel operation method of heterogeneous microgrids is proposed, and the operation cost of the MMG system is reduced by flexible electricity-heat-gas transactions between MGs. However, the limited convergence of Dual decomposition[20] makes the nonlinearity of combined heat and power generation(CHP) not considered. ATC has a stronger convergence than Dual decomposition, but the computation structure relies on the distribution-feedback mechanism between the main system and the subsystems for optimization, and the amount of computation is huge, which is not suitable for real-time scheduling[22]. ADMM is currently the most common method. compared with the previous two algorithms, and which has better convergence ability. In Ref.[23], by combining ADMM and Nash bargaining, the collaborative optimization of MMG and shared energy storage is realized by exchanging Lagrange multipliers. In Ref. [24], ADMM is introduced into bilateral energy transactions, and the conflict of interest of individual participants is solved while ensuring the reliable operation of the system. But, some scholars believe that accurate models need to be built before ADMM [25], but, some nonlinear conditions are not convenient to describe with accurate mathematical models. and Ref. [26] points out that ADMM over-idealizing conditions are unrealistic. In addition, Nash equilibrium provides a new way to solve the problem of stable operation of MMG. Ref. [27] applies Nash equilibrium to energy storage systems and proposes an equilibrium allocation strategy under the premise of considering economic benefits and reliability. In [28], Nash equilibrium is used to protect the privacy of auction participants in combined heat and power markets. In [29], Nash equilibrium is used to solve the problem of multi-energy management of pelagic islanded microgrid clusters. However, Nash equilibrium has a very important premise: if other players do not change strategies, each individual cannot improve their own interests by unilaterally changing strategies [30], which means that it will have a lack of stability in dynamic games. In addition, Nash equilibrium belongs to the category of non-cooperative games, so the contradiction between individual interests and overall interests has always existed [31].

There is no doubt that previous research has greatly promoted the application of MGs and MMG systems. However, the traditional model-based method has been difficult to completely solve high-dimensional, nonlinear, and nonconvex problems in MMG [32-34]. Based on the above analysis, the drawbacks of these methods are listed in Table.1.

Tabel.1 Analysis of the methods mentioned above

| Method | Peculiarity | Drawbacks |
| --- | --- | --- |
| Dual decomposition | Parallel optimization | 1. Central coordinator is not suitable for the distributed framework. 2. Poor convergence ability. 3. Not suitable for nonlinear modelling and real-time scheduling. |
| ATC | Global convergence | 1. A large amount of computation. 2. Not suitable for nonlinear modelling and real-time scheduling. |
| ADMM | Privacy protection | 1. Over-idealizing conditions. 2. Not suitable for nonlinear modelling and real-time scheduling. |
| Nash equilibrium | Better theoretical basis | 1. Hard to reach a global equilibrium. 2. Relying on other algorithms or solvers. 3. Can not effectively solve the partially observable game. |

With the development of artificial intelligence, a kind of model-free algorithm that combines deep learning (DL) and reinforcement learning (RL)–deep reinforcement learning (DRL) is proposed. DRL can sense complex conditions in power systems through deep neural networks and translate OEM problems into the Markov decision process (MDP). This provides a theoretical solution to the drawbacks of model-based methods.

In [35], based on the Deep Q Network (DQN), the author proposes a management strategy for battery energy storage systems and the result shows that deep neural networks can well fit the nonlinear conditions of MGs. Deep Q-learning is the earliest DRL algorithm applied to OEM, which has very important significance. However, it's output interval is discrete, which cannot meet the requirements of practical applications. After applying deep neural networks to the actor-critic framework [36], this problem is solved. The most classic algorithm is the deep deterministic policy gradient (DDPG) [37], which uses the output of the critic to update the actor through gradient ascent. Because the principle is easy to understand and the implementation is simple, DDPG has become one of the most widely used DRL algorithms. Through the comparative test, it is confirmed that the optimization effect of DDPG is better than that of DQN [38]. After improvement, DDPG is used for OEM in partially observable states, which proves that the DRL algorithms can deal with the OEM problem in a complex situation[39]. To optimize the gradient of the actor and control the update speed of the policy, Proximal Policy Optimization (PPO) is proposed [40]. Ref. [41] points out that compared with DDPG, PPO does not depend on the adjustment of parameters and is more applicable. However, Ref. [42] believes that on-policy algorithms are not suitable for scheduling problems of power systems. Several papers [42-43] have confirmed that the application of SAC[44] to power systems is superior to PPO and DDPG.

Due to privacy protection, part of the information in MMG is not shared. The proposal of centralized training with decentralized execution(CTDE)[45] makes the DRL algorithm suitable for the mixed cooperative-competitive environments of MM[46-47].

*1.3. Statement of contributions*

Based on existing research, this paper proposes some new perspectives on the application of MADRL to the OEM problem of MMG. The contributions are summarized as follows:

(1). This paper proposes an energy management-energy auction integrated decision-making framework for MMG systems with SES, which provides a new solution for the management of MMG under the background of energy marketization.

(2). To solve the problem that is difficult to describe the nonlinear conditions of the equipment in MMG. A novel Mixed-Attention is proposed to fit the nonlinear conditions of CHP, BIGCC, CCPP and SES.

(3). Model-based methods are difficult to solve the partially observable dynamic stochastic game in MMG, and can not address real-time scheduling. Therefore, this paper adopts the concept of knowledge fusion, and combines MA-SAC, and MA-WOLF-PHC algorithms to provide a more reasonable operation strategy for the MMG system based on (1) and (2)

(4). To further highlight the advantages of the proposed algorithm, the algorithm in (3) is compared with the mature multi-agent algorithms such as MA-TD3, MA-DDPG, MA-PPO and Nash-Q, and the conclusion is verified from the perspectives of convergence speed and optimization ability.

**2 Models of Multi-microgrids system's scheduling strategy**

This section gives a detailed description of the MMG system (Fig. 1). Taking into account the exchange and bidding of energy between MGs, the mathematical model of the optimization problem is constructed.

*2.1 The framework of the MMG system*

The system consists of three MGs (MG1, MG2 and MG3) with different types of equipment. To meet the demand for large heat load, a CHP unit is applied in MG1, which can use the exhaust heat produced by gas turbines (GT) to supply heat load, and Organic Rankine Cycle (ORC) is used for the sake of improving energy efficiency. Unlike MG1, MG2 utilizes a biomass-based integrated gasification combined cycle (BIGCC) system to solve electricity supply needs, and MG3 uses a combined cycle power plant (CCPP) to provide electricity.

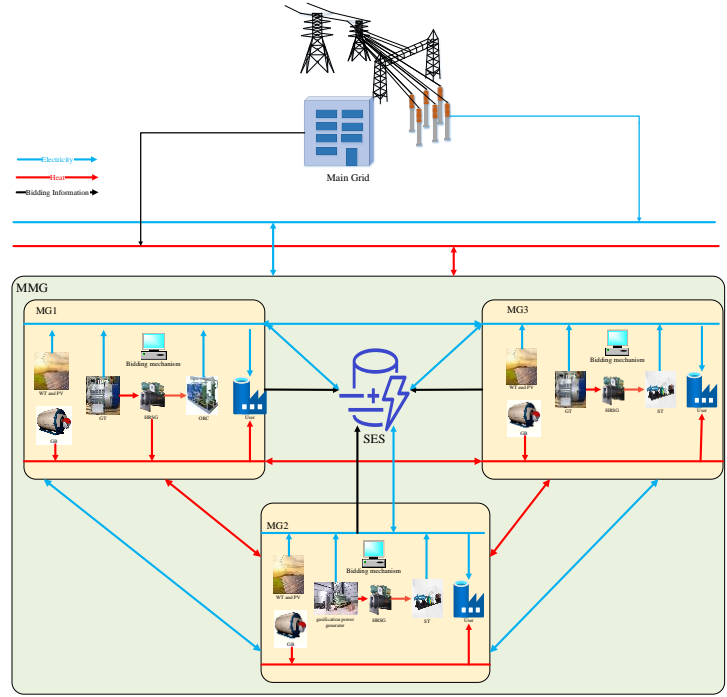

**Fig. 1**. Framework of the MMG system

*2.2 Nonlinear conditions of different types of equipment*

*2.2.1 Photovoltaic array*

The actual electricity generated by the photovoltaic arrays *PV* can be expressed as Eq.(1)[48]. $\eta_{PV}$ is the conversion efficiency, and *S, I,* and $t_0$ respectively represent the array area, solar radiation intensity, and atmospheric temperature.

$$PV = \eta_{PV} SI(1 - 0.005(t_0 + 25)) \tag{1}$$

*2.2.2 Wind turbine*

The output of wind turbines WT is related to air density ρ, blade area A, coefficient factor, $\eta_{WT}$ and wind speed V[49]. The model of which is expressed by Eq.(2).

$$WT = 0.5 \rho A \eta_{WT} V^3 \tag{2}$$

*2.2.3 Gas turbine*

Capstone C200 (C200) is wildly used due to its combinable characteristics and good reliability. Usually, multiple Capstone C200s can be combined, and the capacity is represented by 200·N, the model of which[50] can be expressed as Eq.(3)-Eq.(5).

$$\eta_{C200 \cdot N} = \sum_{i=1}^{k} a_i \left(\frac{P}{200 \cdot N}\right)^{k-i} \tag{3}$$

$$P^{GT} = F_{fuel} \cdot \eta_{C200 \cdot N} \tag{4}$$

$$H(P^{GT}) = \sum_{i=1}^{k} b_k (P^{GT})^i \tag{5}$$

The parameter $\eta_{C200 \cdot N}$ represents the efficiency of C200·N, $P$ denotes the actual load, and $a_i$ is the polynomial coefficient of the output function. $F_{fuel}$ is the heat of fuel. $H(P^{GT})$ is the heat of GT's exhaust gas, which has a nonlinear relationship with $P^{GT}$.

*2.2.4 Heat recovery steam generator*

The heat recovery steam generator (HRSG) is important for the efficient utilization of heat produced by GT. $H(P_{GT})$ passes the superheater, evaporator, economizer in turn. The heat of exhaust gas is collected. In Eq. (6) and Eq. (7), $\eta_{HR}$ is the practical efficiency of HRSG and $H^{HR}$ is the actual output of HRSG[51].

$$\frac{\eta_{HR}}{\eta_{HRN}} = \sum_{i=1}^{n} K_i \cdot \left(\frac{H(P^{GT})}{H^{HRN}}\right)^i \tag{6}$$

$$H^{HR} = H(P^{GT}) \cdot \eta_{HR} \tag{7}$$

*2.2.5 Combined heat and power unit*

In this paper, the CHP consists of a GT, ORC equipment, and HRSG. The output of CHP can be adjusted by controlling the distribution ratio $\beta$, and the model is shown as follows.

$$P^{CHP} = P^{GT} + \beta \cdot H(P^{GT}) \cdot \eta_{ORC} \tag{8}$$

$$H(P^{CHP}) = (1-\beta) \cdot H(P^{GT}) \tag{9}$$

$$0 \leq \beta \leq 1 \tag{10}$$

Due to the function of the ORC, part of the $H(P_{GT})$ can convert into additional electricity. Which is related to the efficiency factor $\eta_{ORC}$, according to Ref.[52]-[53], which is set as 0.1. The electrical output of CHP ($P_{CHP}$) is composed of the electricity generated by ORC and Capstone C200·N. The thermal output of CHP ($H_{CHP}$) is determined by the heat output of C200·N and the heat consumed by ORC.

*2.2.6 Biomass integrated gasification combined cycle system*

Biomass integrated gasification combined (BIGCC) system is mainly composed of the following parts: gasification module, cleaning module, GT, HRSG and steam turbine, the framework of which is shown in Fig. 2.
).

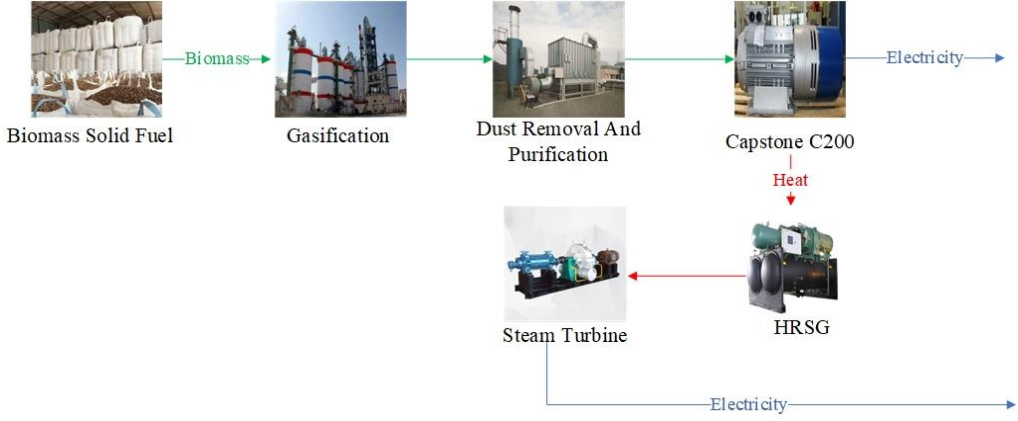

**Fig. 2.** Process of Bigcc system

$$\eta_{BIGCC} = F(\eta_G, \eta_{ST}, \eta_{HR}, \eta_{C200 \cdot N}) \tag{11}$$

The total efficiency of the BIGCC system $\eta_{BIGCC}$ can be expressed as Eq. (11). $\eta_G$ is the efficiency of the biomass gasification module, in this paper, it is set as 76.76%[54], and $\eta_G$ is the efficiency of the GT, which is set as 0.372[55].

*2.2.6 Shared energy storage*

Lithium batteries are widely used due to their green environmental protection and long service life. However, when the storage equipment losses lithium ions, which will lead to the attenuation of capacity. To restore this characterstic as much as possible, the capacity retention is expressed as Eq. (12) in this paper.

$$CR(T) = CR(0) \cdot F(T) \tag{12}$$

The capacity retention $CR(T)$ is affected by initial capacity $CR(0)$ and the decay factor $F(T)$. The parameter $T$ is the cycling times of charge and discharge.

*2.3 Models of the proposed question and restrictions*

In this paper, three MGs can trade energy with each other or with the main grid. The price of energy is time-varying in this paper: $S_t^{main,e}$ and $B_t^{main,e}$ represent the prices of selling electricity to the main grid or buying electricity from the main grid. Correspondingly, $S_t^{main,h}$ and $B_t^{main,h}$ are the transaction prices of heat. $S_t^{micro,e}$, $B_t^{micro,e}$ are the prices of electricity transactions between MGs. Similarly, $S_t^{micro,h}$, $B_t^{micro,h}$ are the prices of heat transactions between MGs.

$$\begin{cases} COST_{1,t}^{MG1} = C_{GS} \cdot [(\frac{P_t^{MG1,GT}}{\eta_{C200 \cdot N_1,t}}) + \frac{H_t^{MG1,GB}}{\eta_{GB,t}}] \\ COST_{1,t}^{MG2} = C_{GS} \cdot \frac{H_t^{MG2,GB}}{\eta_{GB,t}} + C_{BIOMASS} \cdot \frac{P_t^{MG2,GT}}{\eta_{C200 \cdot N_2,t}} \\ COST_{1,t}^{MG3} = C_{GS} \cdot \frac{H_t^{MG3,GB}}{\eta_{GB,t}} + C_{GS} \cdot \frac{P_t^{MG3,GT}}{\eta_{C200 \cdot N_3,t}} \\ 0 \le P_t^{MG1,GT} \le 200 \cdot N_1 (kW) \\ 0 \le P_t^{MG2,GT} \le 200 \cdot N_2 (kW) \\ 0 \le P_t^{MG3,GT} \le 200 \cdot N_3 (kW) \\ 0 \le H_t^{MG1,GB} \le H_{max}^{MG1,GB} \\ 0 \le H_t^{MG2,GB} \le H_{max}^{MG2,GB} \\ 0 \le H_t^{MG3,GB} \le H_{max}^{MG3,GB} \end{cases} \tag{13}$$

Eq. (13) is the cost of equipment in each MG. $C_{GS}$ and $C_{BIOMASS}$ are the prices of natural gas and solid biomass fuel.

$N_i$ is the number of Capstone C200 in MG$i$. $COST_{1,t}^{MGi}$ is the costs of equipment in MG$i$ at time $t$.

$$\begin{cases} COST_{2,t}^{MG1} = \sum_{i=2,3} -S_t^{micro,e} \cdot P_t^{1i,e} + B_t^{micro,e} \cdot P_t^{i1,e} - S_t^{micro,h} \cdot H_t^{1i,h} + B_t^{micro,h} \cdot H_t^{i1,h} \\ COST_{2,t}^{MG2} = \sum_{i=1,3} -S_t^{micro,e} \cdot P_t^{2i,e} + B_t^{micro,e} \cdot P_t^{i2,e} - S_t^{micro,h} \cdot H_t^{2i,h} + B_t^{micro,h} \cdot H_t^{i2,h} \\ COST_{2,t}^{MG3} = \sum_{i=1,2} -S_t^{micro,e} \cdot P_t^{3i,e} + B_t^{micro,e} \cdot P_t^{i3,e} - S_t^{micro,h} \cdot H_t^{3i,h} + B_t^{micro,h} \cdot H_t^{i3,h} \\ P_t^{ij,e} \cdot P_t^{ji,e} = 0 \\ H_t^{ij,h} \cdot H_t^{ji,h} = 0 \\ |P_t^{ij,e}| \leq P^{e,\max}, |P_t^{ji,e}| \leq P^{e,\max} \\ |H_t^{ij,h}| \leq H^{h,\max}, |H_t^{ji,h}| \leq H^{h,\max} \\ \forall i,j \in (1,2,3) \ and \ i \neq j \end{cases} \quad (14)$$

In Eq. (14), $COST_{2,t}^{MGi}$ is the energy exchange cost of $MGi$, which is mainly determined by the amount of traded electricity $P_t^{ij,e}$, and traded heat $H_t^{ij,h}$. The superscript $ij$ means the power flow is from $MGi$ to $MGj$.

After entering the second stage, MGs need to obtain the right to use SES by bidding. The bidding method in this paper is derived from the Sealed-bid Auction. The process can be seen in Fig. 3.

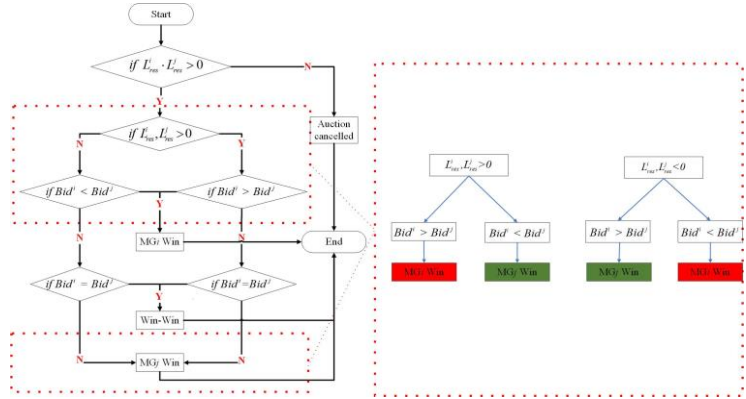

**Fig. 3**. Auction process

Notably, this paper does not completely determine the bidding rules according to the sealed auction method, but makes two changes to it according to the characteristics of the OEM problem:

(1). For keeping the interests of SES and MGs balanced, this paper stipulates that each MG's bid shall not exceed [$S_t^{main,e}$, $B_t^{main,e}$].

(2). To improve the decision-making speed of the agents and avoid a long bidding process, this paper defaults that when the bids of MGs are the same, all are the winning bidders.

The cost of bidding is mainly determined by the bids of MG$i$ ($Bid_i$) and the electricity ($L_{res}^i$) participating in the bidding, which can be expressed as Eq. (15):

$$COST_{3,t}^{MGi} = \begin{cases} Bid_i \cdot Load_i & if MGi \ Win \\ 0 & else \end{cases} \quad (15)$$

Eq. (16)-Eq. (19) show that SES will determine the charging $P^+_{SS,t}$, and discharging $P^-_{SS,t}$ strategy according to the bidding results, the state of charge $SOC_t$ also changes as it charges or discharges.

$$SOC_t = \begin{cases} SOC_{t-1} + P^+_{SS,t} \cdot \eta_{charge} \\ SOC_{t-1} - P^-_{SS,t} / \eta_{discharge} \end{cases} \quad (16)$$

$$0 \leq |P_{SS,t}| \leq P_{SS,\max} \quad (17)$$

$$P^+_{SS,t} \cdot P^-_{SS,t} = 0 \quad (18)$$

$$SOC_T^{\min} \leq SOC_t \leq SOC_T^{\max} \tag{19}$$

Notably, the upper and lower limits of *SOC* are dynamically changed since the battery capacity decay during the time *T* of charge and discharge is considered in this paper, the model is expressed as Eq. (20), and the cost of SES can be represented by Eq. (21).

$$\begin{cases} SOC_T^{\max} = CR(T) \cdot SOC_0^{\max} \\ SOC_T^{\min} = CR(T) \cdot SOC_0^{\min} \end{cases} \tag{20}$$

$$COST_t^{SES} = \begin{cases} -P_{SS,t}^- \cdot Bid + \max(0, (Load_t - P_{SS,t}^-)) \cdot B_t^{main,e} & \text{if } Load > 0 \\ P_{SS,t}^+ \cdot Bid + \min(0, (Load_t + P_{SS,t}^+)) \cdot S_t^{main,e} & \text{if } Load < 0 \\ 0 & \text{else} \end{cases} \tag{21}$$

At the third stage, the losers in biding need to trade the electricity with the main grid to meet the requirements of the energy balance.

$$\begin{cases} COST_{4,t}^{MGi} = -S_t^{main,e} \cdot P_t^{im,e} + B_t^{main,e} \cdot P_t^{mi,e} - S_t^{main,h} \cdot H_t^{im,h} + B_t^{main,h} \cdot H_t^{mi,h} \\ |P_t^{im,e}| \leq P_{\max}^{im,e}, |P_t^{mi,e}| \leq P_{\max}^{mi,e} \\ |H_t^{im,h}| \leq H_{\max}^{im,h}, |H_t^{mi,h}| \leq H_{\max}^{mi,h} \end{cases} \tag{22}$$

For MG*i*, the OEM problem model can be abstracted as Eq. (23). However, *A* and *C* here are nonlinear and non-convex functions, which is very difficult for commercial solvers and model-based methods. Therefore, in the third section of this paper will solve this problem through the MDP.

$$\min \sum_{j=1}^{4} COST_{j,t}^{MGi}$$
$$s.t. \begin{cases} c(x) \leq 0 \\ c_{eq}(x) = 0 \\ A(x) \leq b \\ A_{eq}(x) = b_{eq} \\ l_b \leq x \leq u_b \end{cases} \tag{23}$$

## 3 Algorithms

*3.1 Markov decision process*

A complete MDP consists of a five-dimensional tuple: $\{S, A, P, r, \gamma\}$, $S$ is the collection of states, $A$ is the collection of actions, $P$ is the state transition probability, $r$ is the reward function, and $\gamma$ is the discount factor. The process of MDP is shown in Fig. 4:

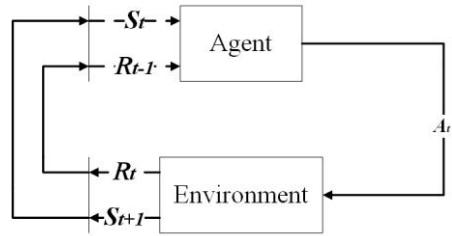

**Fig. 4**. Process of MDP

When the agent receives the information of $S_t$, it outputs $A_t$ according to the policy $\pi(a|s)$. Next, it will receive a reward $R_t$ based on the reward function $r(s,a)$. To evaluation the quality of $\pi(a|s)$, the action value function $Q^\pi(s,a)$ is introduced.

$$Q^\pi(s,a) = E_\pi[G_t \mid S_t = s, A_t = a] \qquad (24)$$

$$G_t = \sum_{k=0}^{\infty} \gamma^k R_{t+k} \qquad (25)$$

According to Eq. (24) and Eq. (25), two functions must be fitted to solve the MDP problem: $\pi(a|s)$ and $Q^\pi(s,a)$. When faced with high-dimensional problems, it is almost impossible to construct accurate functions. But, DRL provides a new solution to these problems.

*3.2 Multi-agent Soft Actor-Critic*

SAC is an algorithm of the Actor-Critic (A-C) framework, which uses the critic network to fit $Q^\pi(s,a)$, and the actor network is used to represent $\pi(a|s)$. By taking the concept of entropy into the A-C framework, not only inherits the data utilization advantage of the off-policy strategy but also improves the degree of active exploration, and the model is expressed by Eq. (26), and the optimal policy can be expressed by Eq. (27).

$$H(x) = E_{x \sim p}(-\log p(x)) \qquad (26)$$

$$\pi^* = \arg\max_\pi E_\pi[\sum_t r(s_t, a_t) + aH(\pi(\cdot | s_t))] \qquad (27)$$

This is a multi-objective optimization problem, and the parameter *alpha* is used to control the importance of action exploitability to the whole objective function, and which can be updated by using a parameterized method and the model is expressed by Eq. (28).

$$Loss(alpha) = E_{s_t \sim s, a_t \sim \pi(\cdot|s_t)}[-alpha \cdot \log \pi(a_t | s_t) - alpha \cdot H_0] \qquad (28)$$

To continuously improve the fitting accuracy of the $Q^\pi(s,a)$, in practical applications, Eq. (29) is used to update the parameter $w_i$ of the critic $i$.

$$Loss(w_i) = E_{(s_t,a_t,r_t,s_{t+1}) \sim s, a_{t+1} \sim \pi_\theta(\cdot|s_t)}[\frac{1}{2}(Q^\pi_{w_i}(s_t,a_t) - (r_t + \gamma(\min_{j=1,2} Q^\pi_{w_{j^-}}(s_{t+1}, a_{t+1}) - alpha \cdot \log \pi(a_{t+1}|s_{t+1}))))^2] \qquad (29)$$

In SAC, two critic networks ($Q_{w1}^\pi$, $Q_{w2}^\pi$) are used to solve the problem that the overestimation of the $Q^\pi(s,a)$. Two target networks ($Q_{w1\text{-}}^\pi$, $Q_{w2\text{-}}^\pi$) are used to improve the convergence stability of $Q_{w1}^\pi$ and $Q_{w2}^\pi$. By combining the Soft Bellman equation with the reparameterization trick to get the loss function of the $\pi(a|s)$, which is expressed by Eq. (30).

$$Loss(\theta) = E_{s_t \sim s}[a \log(\pi_\theta(f_\theta(s_t) | s_t)) - \min_{j=1,2} Q_{w_j}(s_t, f_\theta(s_t))] \qquad (30)$$

In Eq. (30), $f_\theta(.)$ is the rule of reparameterization, after each sampling, the sample value is needed to multiply the standard deviation and add the mean to get the final result, and $\theta$ represents the parameters of the actor network. In this paper, to consider the privacy protection of different MGs, the SAC is extended by CTDE, and the framwork is shown in Fig. 5.

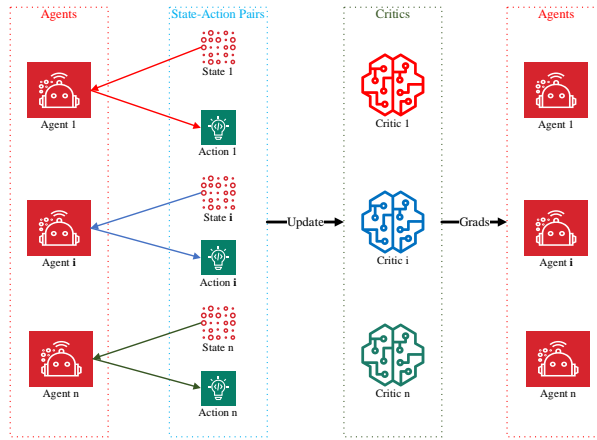

**Fig. 5.** Framework of the CTDE Framework

To combine the CTDE communication mode with the SAC algorithm, the expression for the gradient needs to be re-derived. Suppose that for each agent *i*, denote its action dimension by $a_i \in A$, and $s_i \in S$ is the global state and $\pi$ is the policy, $O_t$ is the gloabel state. When the critic network needs to be updated, at this time, the observed state is the union of the observable variables of all agents. For the critic network of agent *i*, its gradient can be expressed as Eq. (31).

$$Loss(w_{i,k}) = E_{(S_t, A_t, r_{i,t}, S'_t) \square s, a_{i,t} \square \pi_{\theta_{i,t}}(\cdot|s_{i,t})}[\frac{1}{2}(Q^\pi_{w_{i,k}}(O_t, a_{1,t}, a_{2,t}, ..., a_{n,t}) - (r_{i,t} + \gamma(\min_{j=1,2} Q^\pi_{w_{i,j}^-}(O'_t, a'_{1,t}, a'_{2,t}, ..., a'_{n,t}) \qquad (31)$$
$$- a_{i,t} \log \pi_{\theta_i}(a_{i,t} | s'_{i,t}))))^2]_{a'_{i,t} = \pi_{\theta_i}(\cdot|S'_{i,t})}$$

When updating the policy $\pi$, since the critic networks are trained through the global state, and the global state becomes an indirect variable participating in the training process of actor networks. In this training mode, the actor has a certain extension, based on this, Eq. (30) is rewritten as Eq. (32), and the pseudocode of the MA-SAC is shown in algorithm1.

$$Loss(\theta_i) = E_{(S_t, A_t, r_{i,t}, S'_t) \square s}[a_{i,t} \log(\pi_{\theta_i}(f_{\theta_i}(S_{i,t}) | S_{i,t})) - \min_{j=1,2} Q^\pi_{w_{i,j}}(O_t, a_{1,t}, a_{2,t}, ..., f_{\theta_i}(S_{i,t}), ..., a_{n,t})]_{a_{i,t} = \pi_{\theta_i}(\cdot|S_{i,t})} \qquad (32)$$

Algorithm 1

| MA-SAC |
|---|
| Initialize all Critic networks: $Q_{wi,1}^\pi$、 $Q_{wi,2}^\pi$ and Actor network:$\pi_{\theta i}$ |
| Initialize all target networks $Q_{wi,1^-}^\pi$、 $Q_{wi,2^-}^\pi$: |
| $W_{i,1}^- \leftarrow W_{i,1}$, $W_{i,2}^- \leftarrow W_{i,2}$ |
| **For** episode $e=1 \rightarrow E$ **do** |
|   Get the initial global observation state $O$ |
|   **For** time $t=1 \rightarrow T$ **do** |
|     **For** agent $i=1 \rightarrow N$ **do** |
|       Get local observation state $O_t$ |
|       Choose an action $a_{i,t}=\pi_{\theta i}(S_{i,t})$ |
|       Perform $a_{i,t}$ and get the reward $r_{i,t}$ |
|       Obtain next observation state $O_{t+1}$ |
|       Obtain next observation state $S_{i,t+1}$ |
|       Store ($S_{i,t}$, $a_{i,t}$, $r_{i,t}$, $S_{i,t+1}$) in replay buffer $B_i$ |
|       Sample tuples ($S_{i,t}$, $a_{i,t}$, $r_{i,t}$, $S_{i,t+1}$) from $B_i$ |
|       **Train** critic networks by (33) |
|       **Train** actor networks by (34) |
|       **Update** parameter *alpha*$_i$ by (30) |
|       **Update** parameter $W_{1,i}^-$ and $W_{2,i}^-$ by: |
|         $W_{i,1}^- \leftarrow \tau W_{i,1}+(1-\tau) W_{i,1}^-$ |
|         $W_{i,2}^- \leftarrow \tau W_{i,2}+(1-\tau) W_{i,2}^-$ |
|     **End for** |
|   **End for** |
| **End for** |

*3.3 Mixed-Attention*

It can be seen from Eq. (34) that the output of the critic network is an important part of the agent network's gradient, so the critic network is crucial to the performance of the entire algorithm. Therefore, this paper proposes a novel attention mechanism to improve the fitting ability of critic networks. According to this the Mixed Attention Module in Fig. 6 is proposed. In this structure, both the convolution layer and the fully connected layer are used to extract features from the data in the vectors or matries, and the processed data is expanded in parallel and expanded into three-dimensional arrays. Then, the original data will enter the convolutional layer together with the processed data, and get the output through the sigmoid activation function.

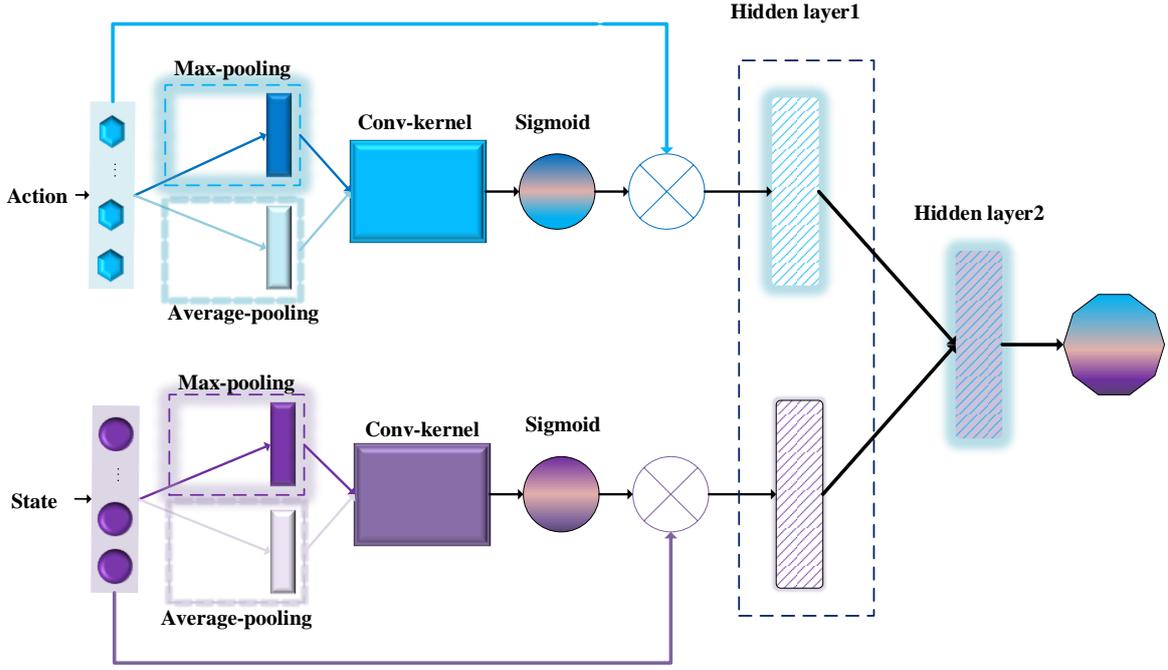

**Fig. 6**. The structure of Mixed-Attention

The structure of the improved critic is shown in Fig. 7.

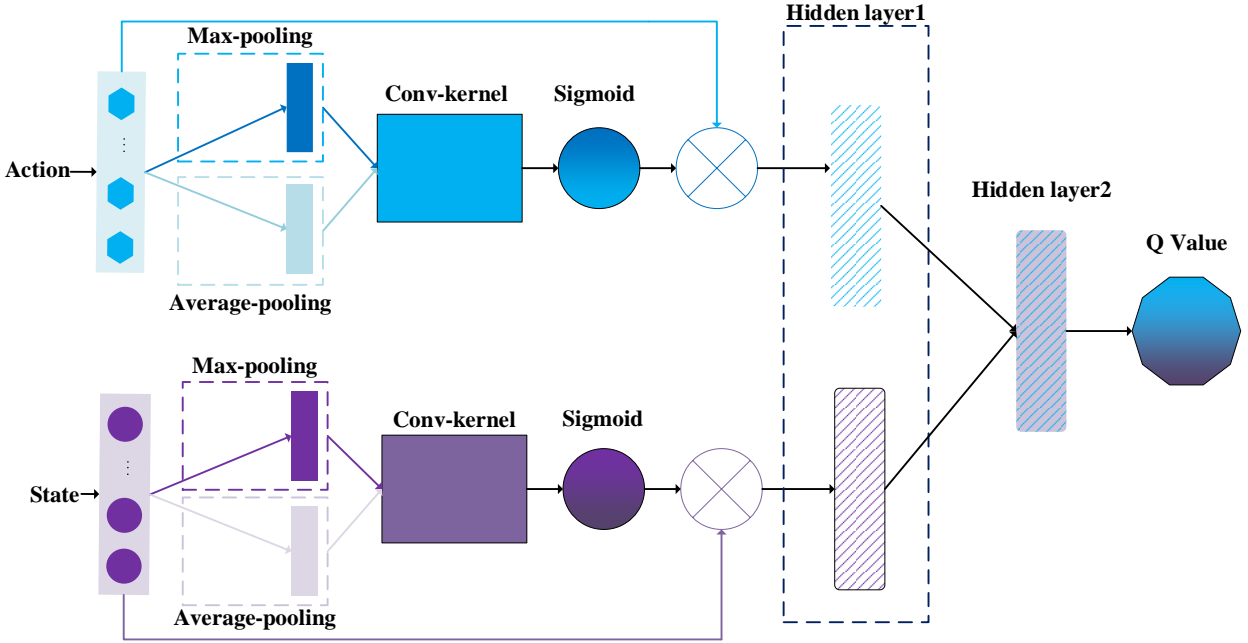

**Fig. 7**. The structure of improved Critic

*3.4 Multi-agent Win or Learn Fast Policy Hill-Climbing*

    In the bidding problem of this paper, since the bidding action is affected by the denomination of the currency, it's range can not be represented by a continuous interval. For this reason, this paper uses CTDE to improve the WoLF-PHC algorithm.

Algorithm 2

| MA-WoLF-PHC |
| --- |
| Initialize $\delta_{a,i}$, $\delta_{w,i}$, $\delta_{l,i}$, let $\delta_{l,i} > \delta_{w,i}$ |
| Initialize $Q_i(S,A) \leftarrow 0$, $\pi_i(s) \leftarrow \frac{1}{|A_i|}$, $C_i(s) \leftarrow 0$, $\pi'_i(s) \leftarrow \pi_i(s)$ |

**For** $i = 1 \to T$ **do**
    **For** agent $i = 1 \to N$ **do**
        Get local observation state $o_i$
        Choose an action $a_i = \pi_i(S_{i,t})$
        Perform $a_i$, and get the reward $r_{i,t}$
        Obtain next observation state $S'_{i,t}$
        Counter the occurrences of $S_{i,t}$ :
$$C(S_{i,t}) = C(S_{i,t}) + 1$$
        Update $Q_i(S, A)$ by:
$$Q_i(S_{i,t}, a_{i,t}) \leftarrow (1 - \delta_a) Q(S_{i,t}, a_{i,t}) + \delta_a (r_i + \gamma \max_a Q(S'_{i,t}, a'_{i,t}))$$
        Update $\pi'_i(s)$ by:
$$\pi'_i(S_{i,t}, a'_{i,t}) \leftarrow \pi'_i(S_{i,t}, a'_{i,t}) + \frac{1}{C(S_{i,t})} (\pi_i(S_{i,t}, a'_{i,t}) - \pi'_i(S_{i,t}, a'_{i,t}))$$
        Update $\pi'_i(s)$ by:
$$\pi_i(S_{i,t}, a_{i,t}) \leftarrow \pi_i(S_{i,t}, a_{i,t}) + \begin{cases} \delta_i & \text{if } a_i = \arg\max_{a'_i} Q_i(S_{i,t}, A'_i) \\ \frac{-\delta_i}{|A_i - 1|} & \text{otherwise} \end{cases}$$
        Calculate $\delta_i$ by:
$$\delta_i = \begin{cases} \delta_{w,i} & \text{if } E(\pi_i) > E(\pi'_i) \\ \delta_{l,i} & \text{otherwise} \end{cases}$$
    **End for**
**End for**

## 4 Knowledge-fusion

To combine the reinforcement learning algorithm with the energy scheduling mechanism, the definitions of state, action, and reward needed to be determined. Fortunately, since key factors such as energy prices and loads are quantifiable, reward and action have physical significance in the context of the question in this paper, these can greatly reduce the difficulty of algorithm application, moreover, because the deep neural network can well fit the nonlinear characteristics of the equipment, it solves the disadvantages of model-based method. Therefore, it is a very good choice to use deep reinforcement learning algorithms to solve the OEM problem of the MMG system. The framework can be seen in Fig. 8.

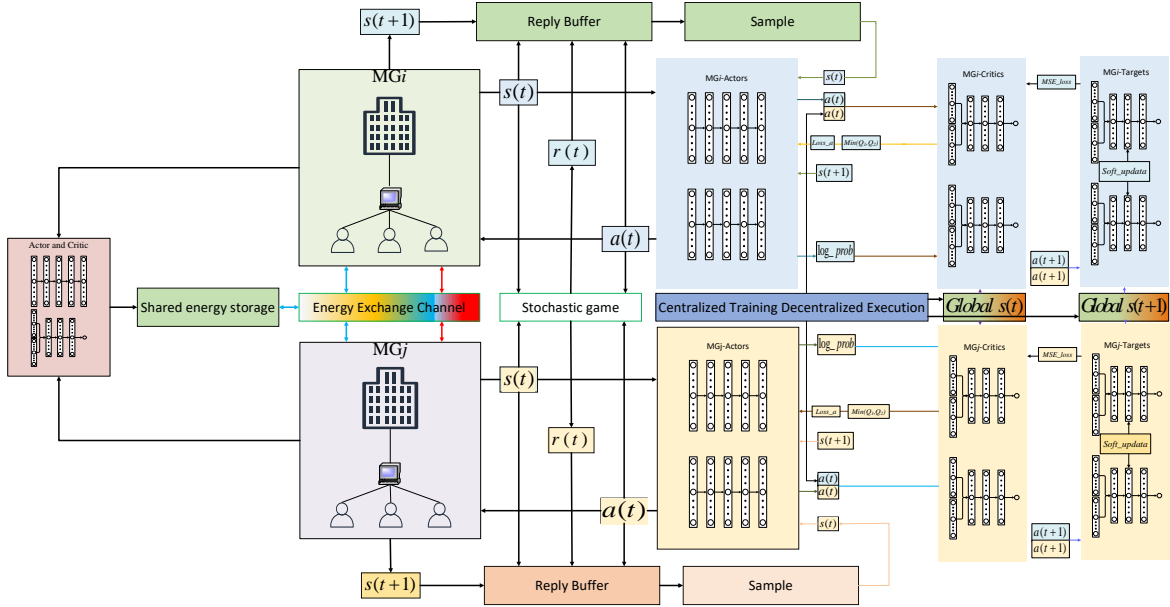

**Fig. 8**. Fusion of MADRL and OEM

*4.1 Physical meaning of state and action*

Ten agents are set up to solve the problem presented in this paper. Among them, each MG will be equipped with three agents, of which two agents are used to manage heat and electricity respectively, and the third agent needs to participate in the bidding to obtain the right to use the SES. The relationship of all agents is shown in Fig. 9.

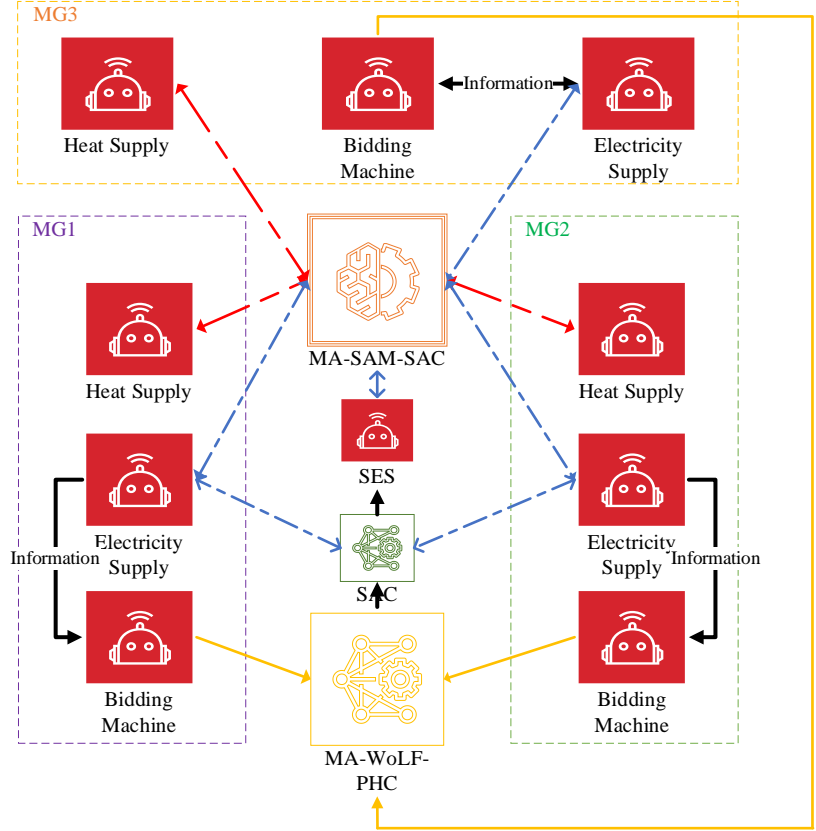

**Fig. 9.** The relationship of all agents

For the electricity supply agent of MG*i*, at time *t*, the observation information includes electrical load, $PV_t$, $WT_t$, and $S_t^{micro,e}$, $B_t^{micro,e}$. Due to the participation of ORC equipment, it is also necessary to observe the heat load to determine the eletrical output of the CHP unit, so the observation of MG1 can be represtented as Eq. (33), and the observation of MG2 and MG3 can be represtented as Eq. (34).

$$S_t^{MG1,e} = \left\{ Load_t^{MG1,e}, Load_t^{MG1,h}, PV_t^{MG1}, WT_t^{MG1}, S_t^{micro,e}, B_t^{micro,e} \right\} \tag{33}$$

$$S_t^{MGi,e} = \left\{ Load_t^{MGi,e}, PV_t^{MGi}, WT_t^{MGi}, S_t^{micro,e}, B_t^{micro,e} \right\}, i \in \{2,3\} \tag{34}$$

For MG*i*'s electricity supply agent, which needs to determine the electrical output of each equipemnt and the electricity traded with each other according to the local observation state. Actions of each MG's electricity supply agent are expressed as Eq. (35)-Eq. (37).

$$A_t^{MG1,e} = \left\{ P_t^{MG1,GT}, P_t^{ORC}, P_t^{12,e}, P_t^{13,e} \right\} \tag{35}$$

$$A_t^{MG2,e} = \left\{ P_t^{MG2,BIGCC}, P_t^{21,e}, P_t^{23,e} \right\} \tag{36}$$

$$A_t^{MG3,e} = \left\{ P_t^{MG2,CCPP}, P_t^{31,e}, P_t^{32,e} \right\} \tag{37}$$

The heat load and $S_t^{micro,h}$, $B_t^{micro,h}$ are necessary to determine the output of the GB in MG$i$. In addition, due to the special characteristic of the CHP unit, for the GB in MG1, it is also required to know the electrical output of the CHP unit and ORC equipment at time $t$. The sate and actions of each MG's heat supply system can be expressed by Eq. (38)-Eq. (42).

$$S_t^{MG1,h} = \left\{ Load_t^{MG1,h}, S_t^{micro,h}, B_t^{micro,h}, P_t^{MG1,GT}, P_t^{ORC} \right\} \tag{38}$$

$$S_t^{MGi,h} = \left\{ Load_t^{MGi,h}, S_t^{micro,h}, B_t^{micro,h} \right\}, i \in \{2,3\} \tag{39}$$

$$A_t^{MG1,h} = \left\{ H_t^{MG1,GB}, H_t^{12,h}, H_t^{13,h} \right\} \tag{40}$$

$$A_t^{MG2,h} = \left\{ H_t^{MG2,GB}, H_t^{21,h}, H_t^{23,h} \right\} \tag{41}$$

$$A_t^{MG3,h} = \left\{ H_t^{MG3,GB}, H_t^{31,h}, H_t^{32,h} \right\} \tag{42}$$

For MG$i$'s bidding machine, the obserable state should be $Load_i$, $S_t^{main,e}$ and $B_t^{main,e}$, and it's action is the bid to the SES, the state and action of bidding machines are shown in Eq. (43)-Eq. (44).

$$S_t^{MGi,b} = \left\{ Load_i, S_t^{main,e}, B_t^{main,e} \right\} \tag{43}$$

$$A_t^{MGi,b} = \left\{ Bid_{i,t} \right\} \tag{44}$$

The state of SES is composed of electrical load $Load_t$ and bid $Bid_t$ of the winning bidders, $SOC_t$, $CR(T)$, $S_t^{micro,e}$, $B_t^{micro,e}$. The actions are the electricity to MG1:$P_t^{s1,e}$, MG2:$P_t^{s2,e}$, and MG3:$P_t^{s3,e}$.

$$S_t^{SES} = \left\{ Load_t, Bid_t, SOC_t, CR(T), S_t^{micro,e}, B_t^{micro,e} \right\} \tag{45}$$

$$A_t^{SES} = \left\{ P_t^{s1,e}, P_t^{s2,e}, P_t^{s3,e} \right\} \tag{46}$$

*4.2 The design of reward functions*

For MMG system, in addition to considering economy, it is also necessary to consider the stability of operation, although a large amount of energy exchange between MGs is beneficial to reduce operating costs, in order to prevent dependencies between MGs, this paper hopes that each MG can cooperate and compete with other MGs while achieving its own independent and stable operation to some extent, considering this situation, an additional penalty term needs to be added to the cost function in section.2.3. In this paper, the unbalance between equipment output and load is added to the cost function as a soft constraint. Eq. (47)-Eq. (49) are the sums of rewards for all equipment in MG$i$, according to the chain rule, each agent will bear the corresponding reward part in Eq. (50).

$$-r_t^{MG1} = \sum_{k=1}^{4} COST_{k,t}^{MG1} + | P_t^{MG1,GT} + P_t^{ORC} - Load_t^{MG1,e} - PV_t^{MG1} - WT_t^{MG1} | + | H_t^{CHP,MG1} + H_t^{MG1,GB} - Load_t^{MG1,h} | \tag{47}$$

$$-r_t^{MG2} = \sum_{k=1}^{4} COST_{k,t}^{MG2} + | P_t^{MG2,BIGCC} - Load_t^{MG2,e} + PV_t^{MG2} + WT_t^{MG2} | + | H_t^{MG2,GB} - Load_t^{MG2,h} | \tag{48}$$

$$-r_t^{MG3} = \sum_{k=1}^{4} COST_{k,t}^{MG3} + | P_t^{MG3,IGCC} - Load_t^{MG3,e} + PV_t^{MG3} + WT_t^{MG3} | + | H_t^{MG3,GB} - Load_t^{MG3,h} | \tag{49}$$

$$\begin{cases} \theta_1' = \theta_1 + \lambda \cdot \dfrac{\partial(r^{MG1}(\theta_1,\theta_2,\theta_3))}{\partial \theta_1} \\ \theta_2' = \theta_2 + \lambda \cdot \dfrac{\partial(r^{MG1}(\theta_1,\theta_2,\theta_3))}{\partial \theta_2} \\ \theta_3' = \theta_3 + \lambda \cdot \dfrac{\partial(r^{MG1}(\theta_1,\theta_2,\theta_3))}{\partial \theta_3} \\ \theta_4' = \theta_4 + \lambda \cdot \dfrac{\partial(r^{MG2}(\theta_4,\theta_5,\theta_6))}{\partial \theta_4} \\ \theta_5' = \theta_5 + \lambda \cdot \dfrac{\partial(r^{MG2}(\theta_4,\theta_5,\theta_6))}{\partial \theta_5} \\ \theta_6' = \theta_6 + \lambda \cdot \dfrac{\partial(r^{MG2}(\theta_4,\theta_5,\theta_6))}{\partial \theta_6} \\ \theta_7' = \theta_7 + \lambda \cdot \dfrac{\partial(r^{MG3}(\theta_7,\theta_8,\theta_9))}{\partial \theta_7} \\ \theta_8' = \theta_8 + \lambda \cdot \dfrac{\partial(r^{MG3}(\theta_7,\theta_8,\theta_9))}{\partial \theta_8} \\ \theta_9' = \theta_9 + \lambda \cdot \dfrac{\partial(r^{MG3}(\theta_7,\theta_8,\theta_9))}{\partial \theta_9} \\ \theta_{10}' = \theta_{10} + \lambda \cdot \dfrac{\partial(r^{SES}(\theta_{10}))}{\partial \theta_{10}} \end{cases} \quad (50)$$

To avoid the over-charged and over-discharged, additional penalties are required when the state of charge is unreasonable, the model can be expressed by Eq. (51), and the coefficient $M$ is the penalty coefficient.

$$r_t^{SES} = COST_t^{SES} + M \cdot \begin{cases} (SOC_{t+1} - SOC_{actual}^{max}) & if\ SOC_{t+1} > SOC_{actual}^{max} \\ (SOC_{actual}^{min} - SOC_{t+1}) & if\ SOC_{t+1} < SOC_{actual}^{min} \end{cases} \quad (51)$$

## 5 Case study

Firstly, Mixed-Attention is used to fit the nonlinear characteristics of equipement in MMG system, and the fitting effects of various attention mechanisms are compared. Secondly, the rationality of the strategy in this paper is verified according to the data of a MMG system in Northwest China. Then, algorithms comparison and scene analysis are taken. Finally, perform a grid search to find the relationship between hyperparameters and the performance of the MA-SAC algorithm.

*5.1 Fitting effects of nonlinear characteristics*

Three indicators: Root Mean Squared Error (RMSE), Coefficient of Determinationard Error ($R^2$), and Mean Absolute Error (MAE) are used to evaluate the fitting effects.

$$RMSE = \sqrt{\frac{1}{N}\sum_{i=1}^{n}(Y_i - f(x_i))^2} \quad (52)$$

$$MAE = \frac{1}{n}\sum_{i=1}^{n}|Y_i - f(x_i)| \quad (53)$$

$$R^2 = 1 - \frac{\sum_{i=1}^{n}(y_i - \hat{y}_i)^2}{\sum_{i=1}^{n}(y_i - \bar{y}_i)^2} \tag{54}$$

The following will verify the effectiveness of Mixed-Attention by comparing the fitting effects of Mixed-Attention-MLP, Spacial-Attention-MLP, Channel-Attention-MLP (SE-NET) and MLP, actual operating data of Capstone C200, data from Ref.[56], visualize the fit effect as Fig. 10.

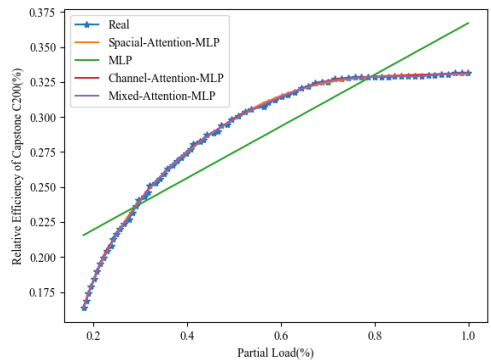

**Fig. 10.** Fitting curves of the Relative Efficiency and Partial Load of Capstone C200

The Capstone C200's power-to-heat ratio (*RH-P*) curve can be seen in Fig. 11. Here, the data used are from Ref. [57].

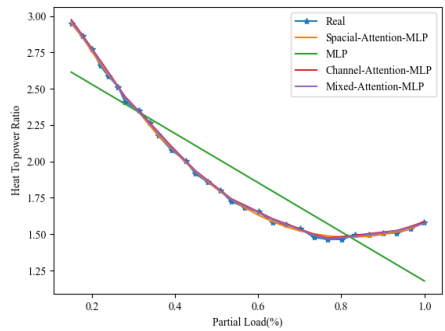

**Fig. 11**. Fitting curves of the Heat-To-Power Ratio and Partial Load of Capstone C200

The curve of the HRSG's efficiency can be seen in Fig. 12. Here, the data used are from Ref. [51].

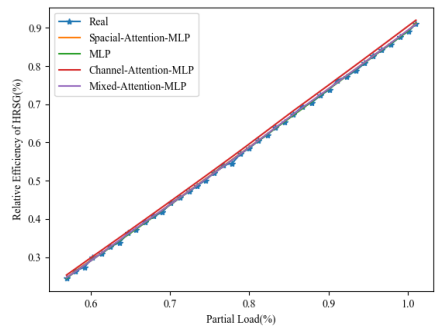

**Fig. 12**. Fitting curves of the Relative Efficiency and Partial Load of HRSG

For the convenience of calculation, in this paper, it is assumed that the rated power of all equipment in the BIGCC system are the same. Therefore, the BIGCC system can be fitted by Eq. (55) and the curve of the efficiency is shown in Fig. 13.

$$\eta_{BIGCC} = \eta_G(\eta_{C200*N} + \eta_{ST}\eta_{HRSG}(R_{H-P} \cdot \eta_{C200*N})) \tag{55}$$

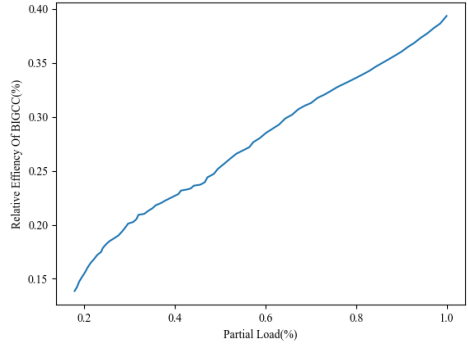

**Fig. 13.** Curve of the Relative Efficiency and Partial Load of BIGCC

The capacity retention of the shared storage with the cycling time at 0.6C charge/discharge rate is shown in Fig.14, and the data are from Ref.[58].

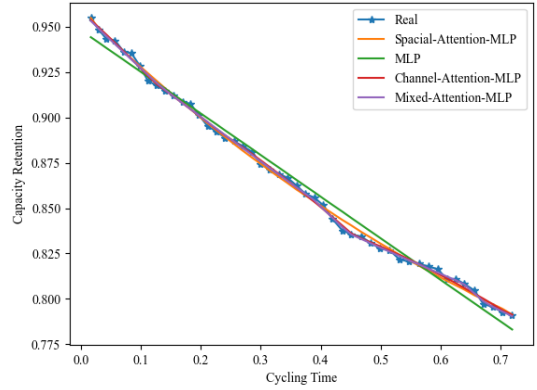

**Fig. 14.** Fitting curves of the Capacity retention and Cycling time of the shared storage

The fitting error analysis is shown in Table 2.

Table 2. Fitting error analysis of nonlinear conditions

| Models | Parameter | RMSE | MAE | $R^2$ |
|---|---|---|---|---|
| Spacial-Attention-MLP | $\eta_{C200}$ | 0.0011 | 0.0009 | 0.9995 |
| | $R_{H-P}$ | 0.0152 | 0.0119 | 0.9989 |
| | $\eta_{HRSG}$ | 0.0032 | 0.0026 | 0.9997 |
| | $CR(T)$ | 0.0023 | 0.0019 | 0.9977 |
| MLP | $\eta_{C200}$ | 0.0211 | 0.0183 | 0.7873 |
| | $R_{H-P}$ | 0.1908 | 0.1660 | 0.8087 |
| | $\eta_{HRSG}$ | 0.0035 | 0.0027 | 0.9997 |
| | $CR(T)$ | 0.0050 | 0.0043 | 0.9889 |
| Channel-Attention-MLP | $\eta_{C200}$ | 0.0008 | 0.0007 | 0.9997 |
| | $R_{H-P}$ | 0.0166 | 0.0129 | 0.9988 |
| | $\eta_{HRSG}$ | 0.0103 | 0.0098 | 0.9973 |
| | $CR(T)$ | 0.0016 | 0.0014 | 0.9989 |
| Mixed-Attention-MLP | $\eta_{C200}$ | 0.0008 | 0.0007 | 0.9998 |

|   |   |   |   |
|---|---|---|---|
| $R_{H\text{-}P}$ | 0.0128 | 0.0105 | 0.9993 |
| $\eta_{HRSG}$ | 0.0032 | 0.0025 | 0.9997 |
| $CR(T)$ | 0.0014 | 0.0012 | 0.9991 |

From the data in Table 2, it can be seen that the higher the degree of nonlinearity of conditions, the better the fitting accuracy of the Mixed-Attention model is than the MLP model. For example, in terms of the fitting effect of $\eta_{C200}$, RMSE, MAE, $R^2$ of Mixed-Attention-MLP are improved by 96.21%, 96.17%, and 26.99% respectively compared with MLP. Even in the case where the nonlinearity is not obvious, such as $CR(T)$, the former can improve the latter by 72.00%, 72.09%, and 1.03%. Therefore, it can be concluded that the combination of the Mixed-Attention can improve the condition fitting ability of MLP.

In the actual application process, it is found that Spatial-Attention needs to expand the data dimension, while Channel-Attention not only needs to expand the data dimension but also needs to change the dimension, therefore, the Spatial-Attention mechanism is more accurate than the channel attention in most cases. The Mixed-Attention does not require additional processing of the data dimension, and retains the physical connection within the data, so the fitting accuracy is the highest.

*5.2 Result analysis of the OEM problem*

Fig. 15, Fig. 16 and Fig. 17 are the results of MMG's OEM problem. As can be seen from the figures, the scheduling strategy proposed in this paper can meet the requirements of stable operation while minimizing the dependence of MMG on the main grid. This not only reduces the burden on the main grid, but also improves the economics of MMG. On the one hand, the MA-SAC algorithm improved by Mixed-Attention can reasonably provide a reasonable energy management strategy for MG1, MG2, and MG3, on the other hand, MA-WoLF-PHC provides a feasible bidding strategy to promote the participation of shared energy storage in the OEM problem by solving the stochastic game.

In Fig. 15(a), Fig. 16(a) and Fig. 17(a), electricity exchange occurs between MG1, MG2 and MG3 several times. Between 10:00-20:00, in MG1, due to the sudden increase in electrical load and the decrease in PV and WT, there is a lack of electricity, and this part of the electric energy just makes up for the electricity imbalance of MG2 and MG3. In this case, both the electricity purchased by MG1 from the main grid and the electricity sold by MG2, MG3 to the main grid can be reduced, which is very helpful to keep the main grid running stably. When considering the participation of SES and energy trading between MGs, which reduce the number of times MG1 exchanges electricity with the main grid from 18 times to 5 times, and the total electricity exchanged is reduced from 843.9kW to 272.1kW in 24h. For MG2, the amount of electricity exchanged with the main grid is reduced from 941.2kW to 109.4kW. For MG3, is 1751.8kW to 920.0kW. It can also be found in Fig. 18(a) that due to the continuous increase of the power load of MG1 during 11:00-20:00, the ORC equipment can also be flexibly controlled.

In Fig. 15(b), Fig. 16(b) and Fig. 17(b), heat exchange also occurs between MG1, MG2 and MG3. In MG1, since the electrical load falls and the thermal load rises between 0:00-5:00, the CHP unit has a lack of heat. Due to the advantages of the multi-agent communication method of CTDE, the GB of MG2 and MG3 can perceive this change, thereby rising their own output to achieve the purpose of providing the heat energy to MG1. And since the electrical load rises and the thermal load falls between 12:00-17:00, MG1 also exchanges excess heat energy to MG2 and MG3.

Finally, it should be noted that considering that this is a stochastic game problem, all three MGs control their output equipment in a relatively conservative manner. Because it can be seen from the figures that their electrical and thermal outputs are close to the actual load. This is reasonable because MGs cannot obtain data from other operators when they are actually running.

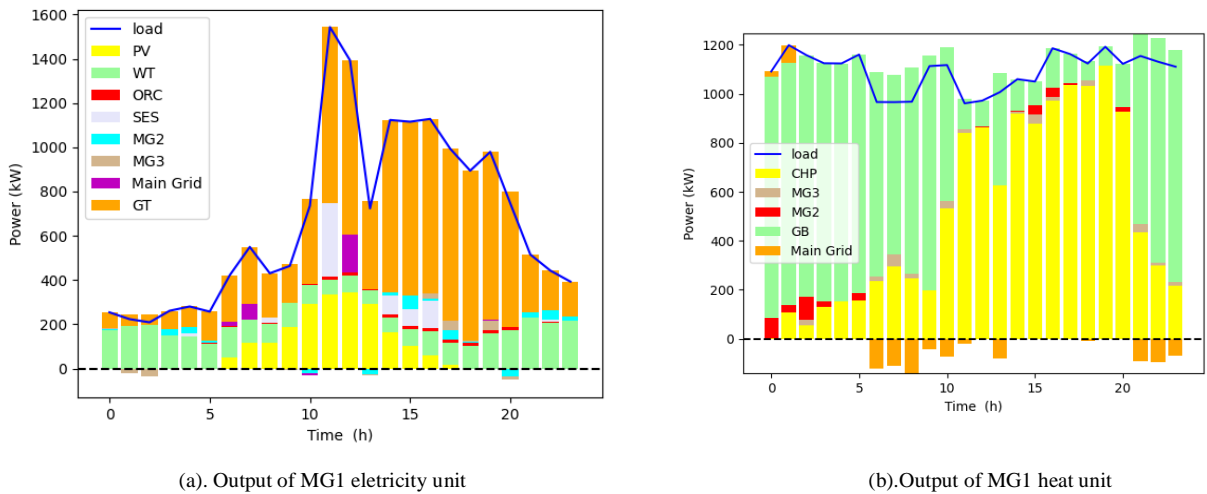

(a). Output of MG1 eletricity unit  (b).Output of MG1 heat unit

**Fig. 15**. OEM results of MG1

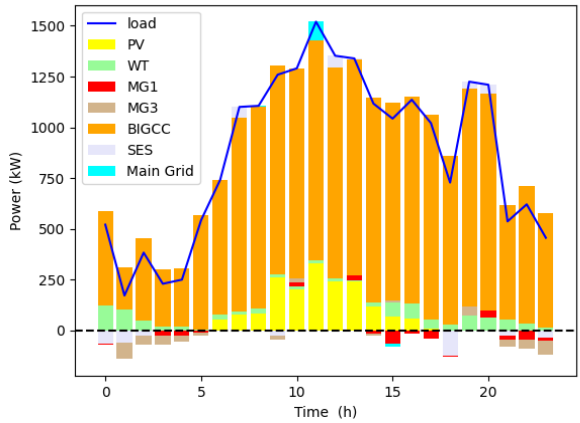

(a). Output of MG2 eletricity unit

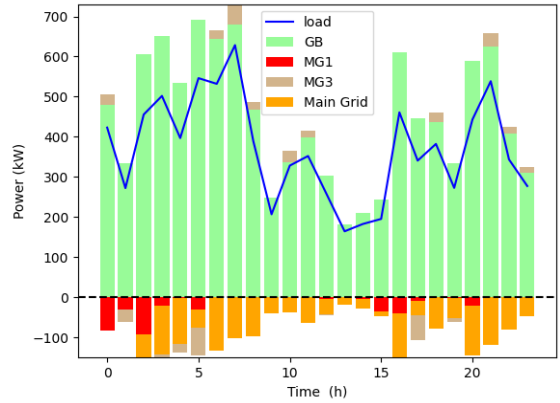

(b).Output of MG2 heat unit

**Fig. 16.** OEM results of MG2

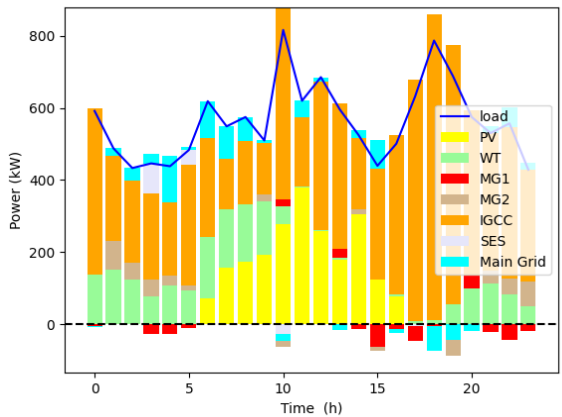

(a). Output of MG3 eletricity unit

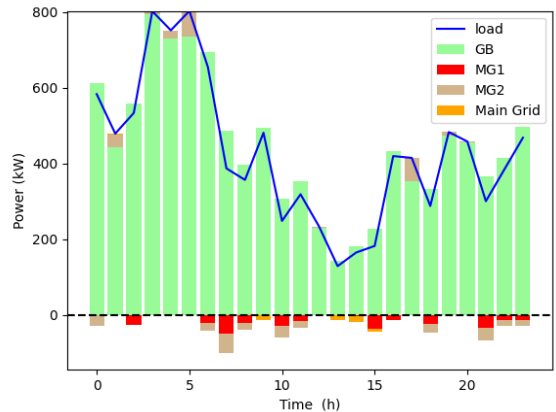

(b).Output of MG3 heat unit

**Fig.17.** OEM results of MG3

Fig. 18 and Fig. 19 show the bidding results of all MGs and the electricity participating in the bidding. In most cases, the bids of the three MGs are between the electricity sales price and electricity purchase price of the main grid. In particularly cases, when there is a large amount of electricity imbalance, for example, at 11:00, MG1 had a very large electricity shortage, so it bids at a high price of 0.0875\$ (The price ceiling is 0.1\$). In other cases, the bidding power of three MGs is relatively close, and MGs will make relatively reasonable bids based on the unbalanced power.

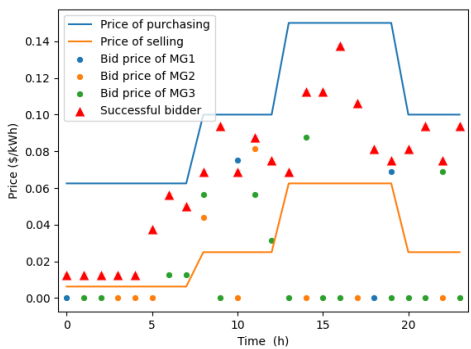

**Fig. 18**. Auction results

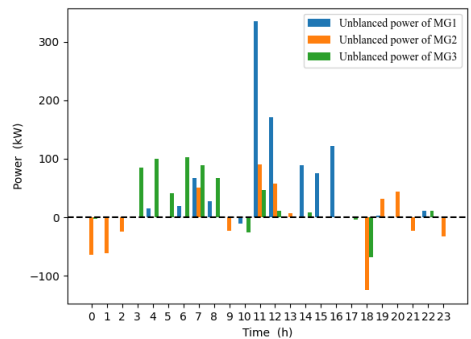

**Fig. 19**. Electricity participates in the auction

Fig. 20 and Fig. 21 are the state change of SES in 24 hours, the results show that SES can actively participate in the energy time-shift of the MMG system under the incentive of bidding activities. With the increase of charge and discharge times, its own capacity retention also exhibits nonlinear decay.

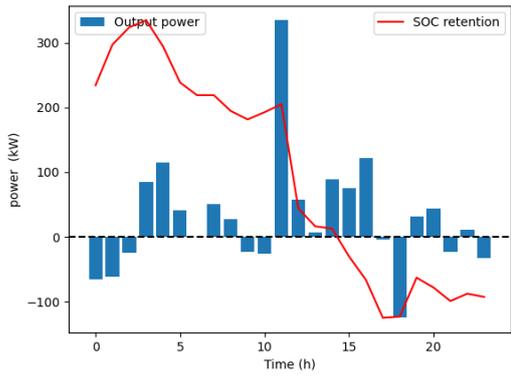
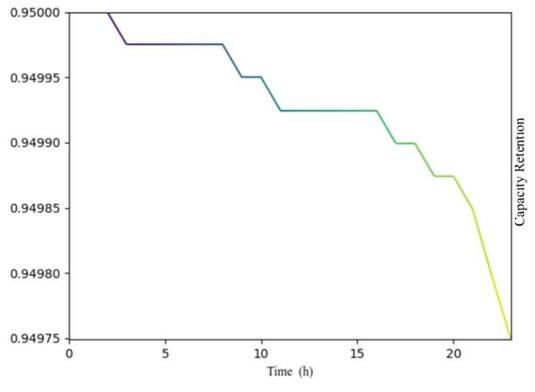

Fig. 20. Power output and SOC of SES          Fig. 21. Capacity retention of SES

Fig. 22 shows the changes in the condition of different types of equipment, according to Fig. 15-Fig. 17, The conditions of the equipment change with the dynamical sates.

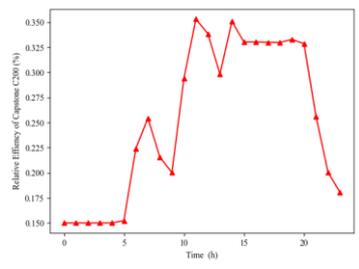
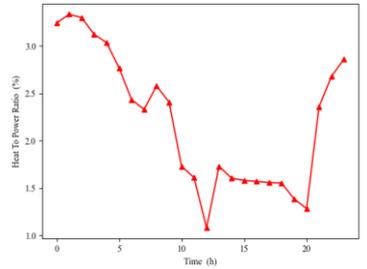
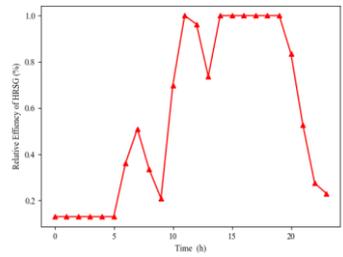

(a). Nonlinear condition of GT in MG1     (b). $R_{H-P}$ of GT in MG1     (c). Nonlinear condition of HRSG in MG1

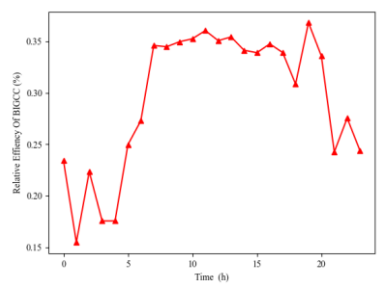

(d). Nonlinear condition of BIGCC in MG2          (e). Nonlinear condition of CCPP in MG3

Fig. 22. Nonlinear conditions of equipment in MGs

In general, the MMG system composed of different types of equipment is complementary. Although the CHP unit of MG1 can greatly improve the utilization efficiency of natural gas, the adjustment of the heat to power ratio lacks flexibility, which can easily lead to energy imbalance, even in some cases, a large amount of imbalance will occur. In contrast to MG2 and MG3, since there is no direct connection between the BIGCC (CCPP) system and the GB, the electricity and heat can be better managed. It is in this case that MG2,MG3 and MG1 have a cooperative relationship to a considerable degree.

The right to use SES often belongs to different MGs at different times, at 11:00 and during 15:00-16:00, SES mainly solves the electricity imbalance problem of MG1, and which is mainly used by MG2 during 0:00-2:00 and 18:00-21:00.

During 3:00-5:00 and 13:00-17:00, MG3 acquired the right to SES. This also shows that each MG has different demand for energy storage, it is difficult to set a reasonable energy storage capacity for different MGs, and the emergence of SES can reasonably solve this problem, In addition, due to the involvement of SES, MGs do not directly trade electrical energy with the main grid, which reduces the burden on the operation of the main grid.

Finally, it can be seen from the results that there is no obvious dependence between three MGs, and the energy exchange only occurs at specific moments, which means the proposed method can not only protect the interests of each operator but also ensure the stable operation of the MMG system.

*5.3 Algorithms and scenes comparison in energy management of MMG*

*5.3.1 Performance of MA-SAC (SAC) algorithm with different attention mechanisms*

This paper sets up six groups of controlled experiments to compare the effects of combining different attention mechanisms (Spatial-Attention, Channel -Attention, Self-Attention, Linear-Attention, and Mixed-Attention) with the MA-SAC or SAC algorithm. In order to avoid the chance of a single test, each version of the MA-SAC (SAC) algorithm is tested three times. The results can be seen in Fig. 23. and the average return of different attention mechanisms can be seen in Table 3.

It should be noted that considering the similar structure of MG2 and MG3, and in order to reduce the redundant content of the paper, in 5.3 and 5.4, the agents of MG1, MG2, SES will be the mainstay, and the convergence curve of the MG3's agents can be seen in the appendix.

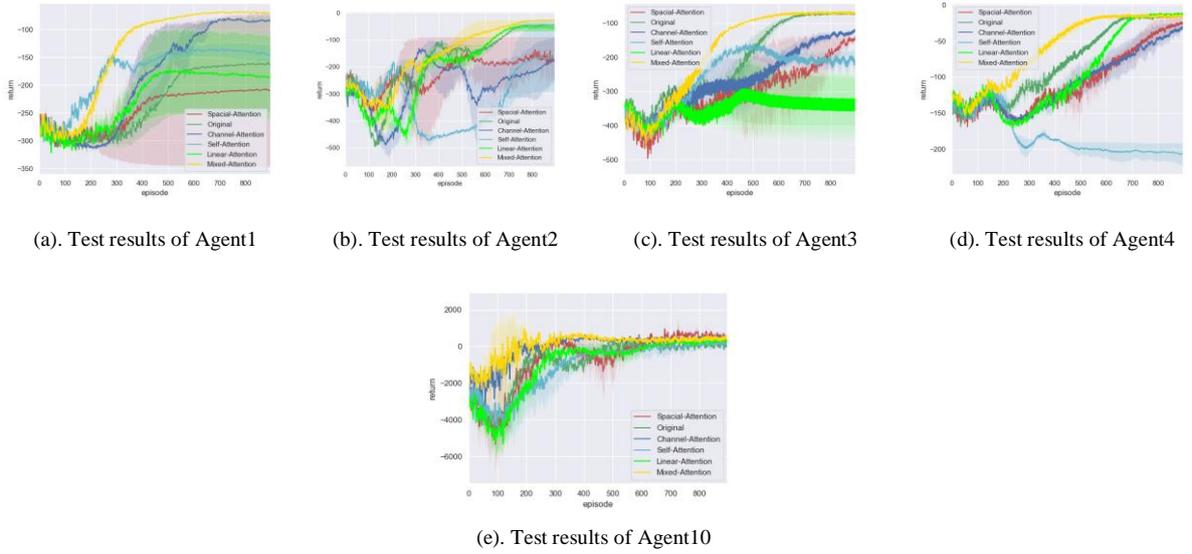

(a). Test results of Agent1  (b). Test results of Agent2  (c). Test results of Agent3  (d). Test results of Agent4

(e). Test results of Agent10

**Fig. 23**. Test results of different attention mechanisms

Table 3. Average return of different attention Mechanisms

| Attention Mechanism | Agent1 | Agent2 | Agent3 | Agent4 | Agent10 |
|---|---|---|---|---|---|
| Mixed-Attention | -136.0 | -142.0 | -178.1 | -57.49 | 258.9 |
| Spatial-Attention | -243.4 | -201.2 | -276.7 | -94.57 | -267.9 |
| Linear-Attention | -223.3 | -203.3 | -347.9 | -93.3 | -325.7 |
| Channel-Attention | -186.6 | -300.9 | -260.2 | -136.6 | -45.83 |
| Self-Attention | -166.5 | -319.7 | -240.2 | -177.2 | -108.8 |
| None | -231.1 | -180.4 | -211.4 | -78.07 | -867.0 |

Agent1-Agent4 are the electricity supply systems and heat supply systems of MG1 and MG2, They are all controlled with MA-SAC algorithm. Agent10 is the SES, which is controlled by the SAC algorithm. In Fig.23(a), Mixed-Attention, Self-Attention and Channel-Attention all help to improve the performance of the original MA-SAC. Unlike agent1, agent2 is affected by the characteristics of the CHP unit. In addition to observing the thermal load of MG1, it also needs to observe the actions of agent1. In Fig.23(b), both Linear-Attention and Mixed-Attention can make MA-SAC converge to reasonable results, but the former slightly degrades the performance of the original MA-SAC.

Agent3 is the electricity supply system of MG2, the critics of whose need to consider a large number of nonlinear conditions, in this case, Mixed-Attention is the best in terms of convergence speed and optimization effect. The average return of the original MA-SAC algorithm in Table.3 is -211.4, and after adding Mixed-Attention, it increases to -178.1, an increase of 15.75%. In Fig.23(d), Mixed-Attention significantly improves the convergence speed of MA-SAC, the optimization effect is slightly worse than Linear-Attention, and the return value after convergence is similar to the original MA-SAC. It should be pointed out that Self-Attention causes MA-SAC to fail to converge.

Agent10 controls the action of the SES. In Fig. 23.(e), the original the original version has the worst result, although Self-Attention has slightly improved the performance of MA-SAC, in curve, it does not make the SES reach a stable profitable state (the return after convergence is less than 0). Channel-Attention and Mixed-Attention have similar effects and are slightly worse than Spatial-Attention, but their convergence speed is fast.

In general, different types of attention mechanisms improve the performance of the MA-SAC algorithm in different agents, but these attention mechanisms do not guarantee that they can improve the MA-SAC algorithm during the training process of each agent's performance. For example, during the training process of agent1 and agent3, Channel-Attention appears to fall into the local optimum, which is also the case with Linear-Attention, the Spatial-attention has a problem of slow convergence, and in the training process of agengt1, very poor results have been obtained. In Fig.23(c), Self-Attention has a very serious non-convergence situation.

*5.3.2 Performance of different MADRL algorithms*

In order to compare the performance of different DRL algorithms in the questions proposed in this paper, this paper tests the effects of MA-SAC, MA-DDPG, MA-TD3 and MA-PPO, and the test results are shown in Fig. 24.

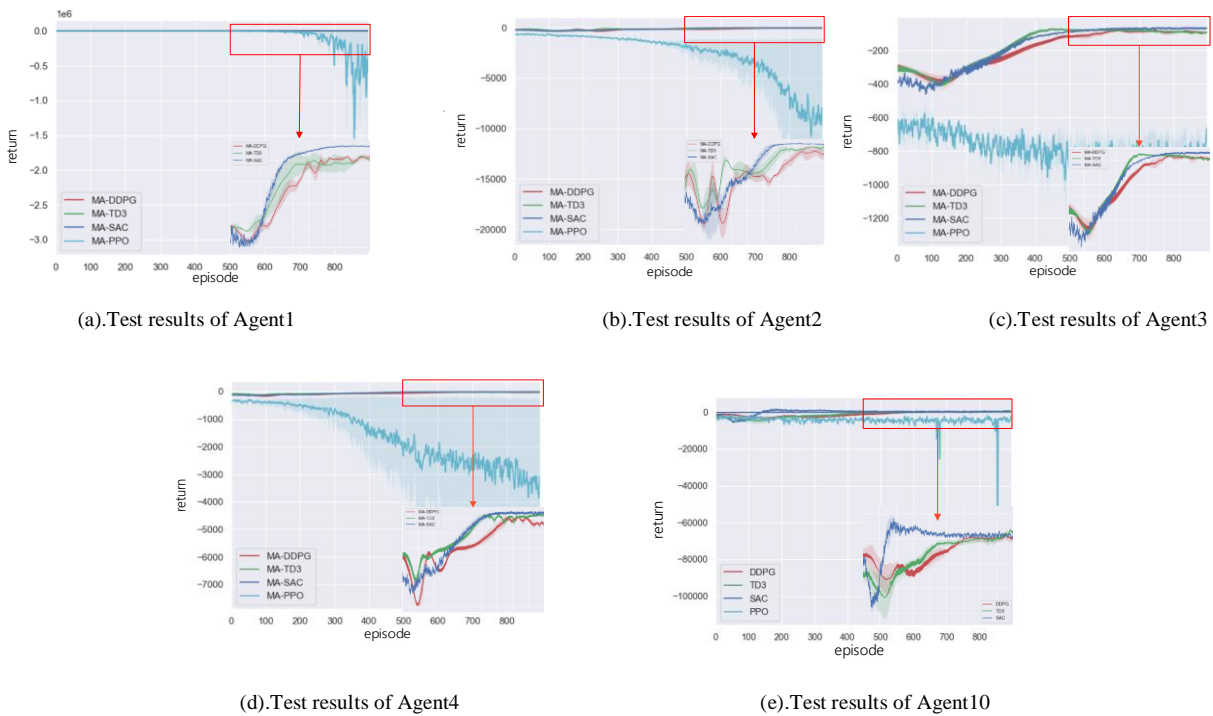

(a).Test results of Agent1  (b).Test results of Agent2  (c).Test results of Agent3

(d).Test results of Agent4  (e).Test results of Agent10

**Fig. 24**. Test results of different MADRL algorithms

For the test results in Fig. 24, it should be noted that in order to ensure fairness, the neural network structure of all algorithms is the same, and the public parameters such as the learning rate, the size of the replay buffer and the minibatch are also the same. In addition, to restore the on-policy characteristics of the MA-PPO algorithm, this paper does not set the replay buffer for the MA-PPO algorithm. In the test of this paper, the effect of MA-PPO is the worst, which is consistent with the view proposed by Ref. [42]. In most cases, the results of MA-TD3 are a little better than MA-DDPG. However, since MA-TD3 has a relatively small exploration space, so, it performs worse than the MA-SAC algorithm in the test. This paper compares the MA-DDPG, MA-SAC, and MA-TD3 algorithms in a 30-day test set, Take the total operating cost of each operator in the MMG system within 30 days as the measurement.

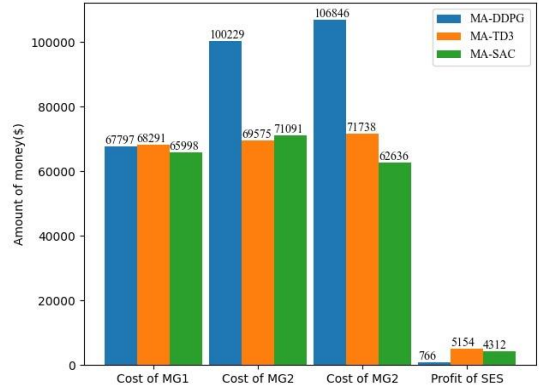

**Fig. 25**. Results of different algorithms on the test set

In Fig. 25, it is obvious that the operating cost of MG1 and MG3 controlled by MA-SAC is the lowest, the SES controlled by TD3 algorithm, whose profit is slightly higher than that of SES controlled by SAC algorithm. However, the profit of SES is not only related to its charging and discharging strategy, but also related to operation policies of MG1, MG2 and MG3. From the perspective of the total operation cost of the entire MMG system, MA-SAC (SAC) is the best algorithm.

The MA-WoLF-PHC algorithm is used to provide bidding strategies to MG1, MG2 and MG3, this part compares it with the Nash Q-Learning (Nash Q) algorithm. The result is shown in Fig. 26.

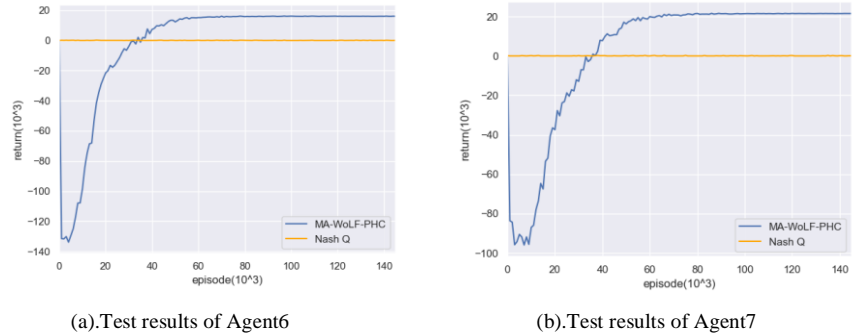

(a).Test results of Agent6                    (b).Test results of Agent7

**Fig. 26**. Test results of MA-WoLF-PHC and Nash Q

The results of Nash Q are very poor. This is because in the multi-agent environments, the policy of each participant is dynamic, and all participants need to actively change their policies according to the environment. Compared to the Nash Q, MA-WoLF-PHC does not pursue accurate solutions, but pursues continuous improvement and optimization of strategies. Therefore, it can also achieve good results in the dynamic environment of stochastic games.

*5.3.3 Scenes Comparison*

To verify the effect of SES and the energy exchange between MGs on improving the economy of MMG system, in this paper, the cost of three scenes in the 30-day test set is calculated. Scene 1: considering SES and allowing energy trading between MGs. Scene 2: without SES, but energy trading is allowed. Scene 3: MGs only have energy trading with the main grid. The results are shown in Fig. 27.

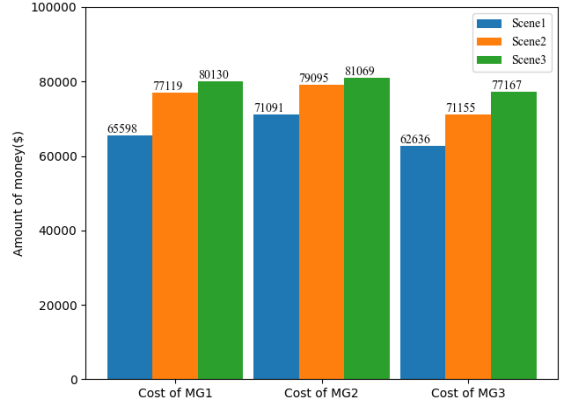

**Fig. 27**. Result of scenes comparison

In Fig. 27, When SES is not involved in the strategy of the MMG system, the cost of MG1 will increase by $11521, the cost of MG2 will increase by $8004, and the cost of MG3 will increase by $8519. On this basis, if the energy trading between MGs is canceled, the cost of the entire MMG system will increase by 39041$. Therefore, the participation of SES and the energy trading mechanism of MGs can contribute to the economic stability of the entire MMG system.

Table 4. Scene comparison

|     | Scenes  | Cost($) | Electricity traded with the main grid(kW) |
| --- | ------- | ------- | ----------------------------------------- |
| MG1 | Scene 1 | 65998   | 22094                                     |
|     | Scene 2 | 77119   | 24634                                     |
|     | Scene 3 | 80130   | 46509                                     |
| MG2 | Scene 1 | 71091   | 12370                                     |
|     | Scene 2 | 79095   | 31805                                     |
|     | Scene 3 | 81069   | 63187                                     |
| MG3 | Scene 1 | 62636   | 15963                                     |
|     | Scene 2 | 71155   | 23080                                     |
|     | Scene 3 | 77167   | 36652                                     |

In Tab 4, It can be clearly seen that when the energy transaction between MGs and the participation of SES is considered, the energy traded between MG1 and the main grid is reduced from 46509 kW to 22094 kW, while MG2 is reduced from 63187 kW to 12370 kW. For MG3, the numbers are 36652 kW and 15963 kW.

*5.4 Hyperparameters Analysis*

Learning rate of alpha, initialization of alpha and target entropy are considered to be the hyperparameters that have the greatest impact on the SAC algorithm, in Ref. [59], the author believes that setting target to a negative number of the agent's action dimension is ideal, therefore, this paper focuses on exploring the impact of the first two hyperparameters on the algorithm.

In this section, grid search is used to explore the effect of several special hyperparameters on the performance of the SAC algorithm, the search results can be seen in Fig. 28.

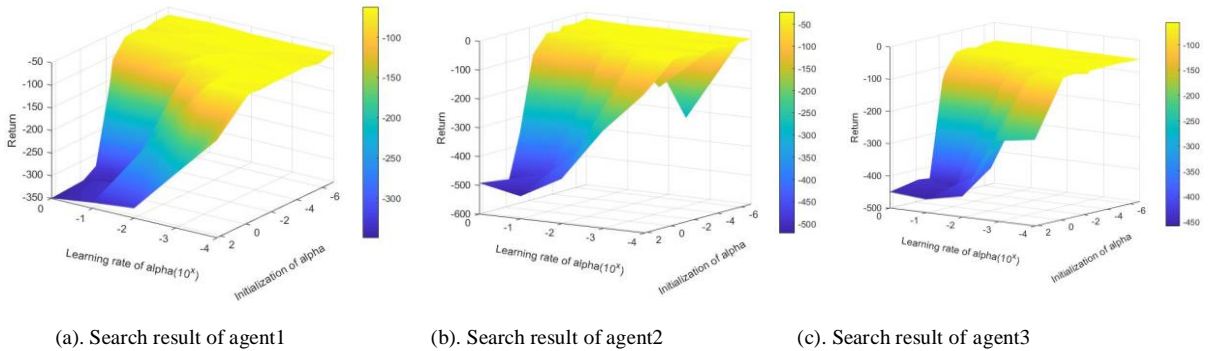

(a). Search result of agent1  (b). Search result of agent2  (c). Search result of agent3

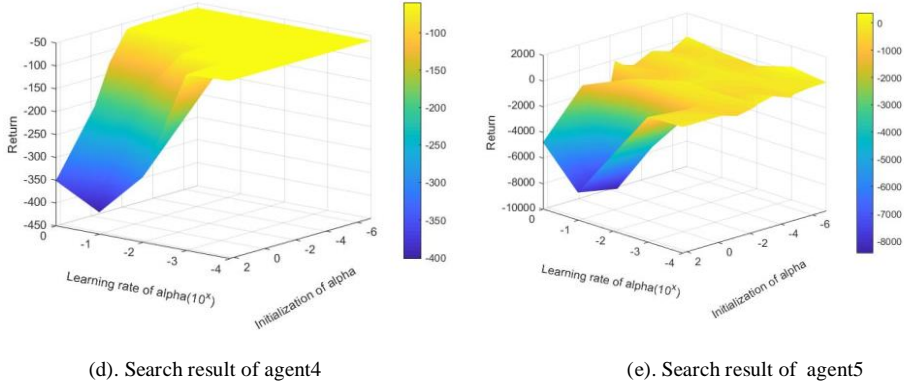

(d). Search result of agent4       (e). Search result of agent5

**Fig. 28**. Relationship between Learning rate of alpha, Initialization of alpha and Return

The 3D surface distribution of the five agents about the two hyperparameters and the return after convergence is consistent, which explains that the search results are reasonable. In addition, it can be known from the Fig. 28 that a larger learning rate of alpha requires a smaller initialization of alpha to ensure the algorithm converges, and when the learning rate is reduced to around $10^{-3}$, this effect is greatly reduced. For the hyperparameter of the initialization of alpha, if it is greater than 0, there will be no convergence, in contrast to the former, larger initialization of alpha tend to require smaller learning rates to ensure convergence, the dependency starts to disappear when the initialization of alpha is around log(0.01).

**6 Conclusions**

Aiming at the solving the stochastic games between MGs with nonlinear conditions, a reasonable strategy is proposed and tested based on the operation data of a MMG in Northwest China, and the following conclusions are drawn:

(1). The Mixed-Attention proposed in this paper can be used to fit the nonlinear conditions of different types of equipment, and can improve the performance of MA-SAC.

(2). The MA-SAC algorithm can effectively solve the OEM problem of MMG considering the nonlinear conditions of equipment. According to the test set results, and the performance is better than MA-DDPG and MA-TD3.

(3). The MA-Wolf-PHC algorithm can solve the problem of MGs bidding. Compared with Nash-Q, it has better convergence effect and better ability to cope with dynamic environment.

(4). According to the results of scene analysis, compared with not considering the energy transaction between microgrids, the MA-SAC algorithm is used to realize the electricity and heat transaction between MGs, which can significantly reduce the operation cost of MMG, and after the participation of SES, the operation cost of MMG is further reduced. In addition, SES can also reduce the energy of MGs trading with the main grid, thus reducing the burden of the main grid operation.

(4) From the results of scene analysis, When considering energy trading between MGs, the MMG operating cost can be reduced by 11599$ in the 30-day test set, and when the SES is involved, the number becomes 38641$.

(5)It is found that a larger Learning rate of alpha and a larger Initialization of alpha are not conducive to the algorithm convergence. When the learning rate of alpha is large, the Initialization of alpha needs to be adjusted to be much smaller than the target alpha, and when the initialization of alpha is greater than log(0), it doesn't converge. However, when the learning rate is reduced to about $10^{-3}$ and the initialization value is reduced to about $\log(10^{-2})$, this problem can be easily alleviated.

It should be pointed out that this paper puts forward a reasonable management strategy for the combined heat and power MMG system, but the cooling load is not considered, In addition, from the perspective of stable operation, this paper does not analyze the impact of accidental scenarios on the operation of MMG systems, These problems will be deeply analyzed in the future.

**Appendix**

Table A1. Hyperparameters of MA-Mixed-Attention-SAC

| Hyperparameters | Value |
| --- | --- |
| Target_alpha | -dim(action) |
| Initilization of alpha | Log(0.01) |
| Learning rate of alpha | 0.003 |
| Learning rate of critics | 0.00005 |
| Learning rate of actors | 0.000005 |
| Hidden units | 64 |

Table A2. Parameters of MA-WoLF-PHC

| Parameters | Value |
|---|---|
| $\gamma$ | 0.8 |
| $\delta_{w,i}$ | 0.05 |
| $\delta_{l,i}$ | 0.1 |

Table A3. Parameters of MMG

| Parameters | Value |
|---|---|
| Capacity of PV in MG1 | 400kW |
| Capacity of WT in MG1 | 300kW |
| Capacity of PV in MG2 | 400kW |
| Capacity of WT in MG2 | 200kW |
| Capacity of PV in MG3 | 400kW |
| Capacity of WT in MG3 | 200kW |
| $N1$ | 4 |
| $N2$ | 6 |
| $N3$ | 5 |
| $CR(0)$ | 3000kW |
| $H_{max}^{MG1,GB}$ | 1000kW |
| $H_{max}^{MG2,GB}$ | 600kW |
| $H_{max}^{MG3,GB}$ | 1000kW |
| $\eta_{Charge}$ | 0.95 |
| $\eta_{discharge}$ | 0.95 |
| $SOC_0^{max}$ | 0.9 |
| $SOC_0^{min}$ | 0.1 |
| $P_{ss,max}$ | 400kW |
| $P_{max}^{im,e} = P_{max}^{mi,e}$ | 400kW |
| $H_{max}^{im,h} = H_{max}^{mi,h}$ | 300kW |

Table A4. Prices for different energy

|  | 0:00-8:00 | 8:00-12:00 | 12:00-19:00 | 19:00-24:00 |
|---|---|---|---|---|
| Biomass | 0.0353$ | 0.0353$ | 0.0353$ | 0.0353$ |
| Gas | 0.0494$ | 0.0494$ | 0.0494$ | 0.0494$ |
| $S_t^{micro,e}$ | 0.0212$ | 0.0565$ | 0.0847$ | 0.0565 |
| $B_t^{micro,e}$ | 0.0424$ | 0.1000$ | 0.1412$ | 0.1000$ |
| $S_t^{main,h}$ | 0.0282$ | 0.0282$ | 0.0282$ | 0.0282$ |
| $B_t^{main,h}$ | 0.1342$ | 0.1342$ | 0.1342$ | 0.1342$ |
| $S_t^{micro,h}$ | 0.0494$ | 0.0494$ | 0.0494$ | 0.0494$ |
| $B_t^{micro,h}$ | 0.1059$ | 0.1059$ | 0.1059$ | 0.1059$ |

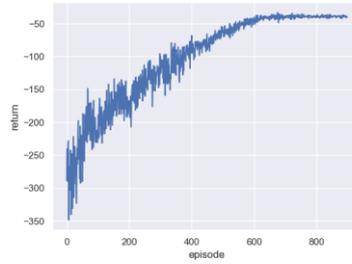

**Fig. A1**. Convergence curve of MG3's electricity supply system

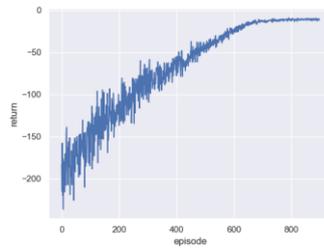

**Fig. A2**. Convergence curve of MG3's heat supply system

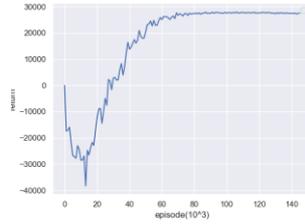

**Fig. A3**. Convergence curve of MG3's bidding machine

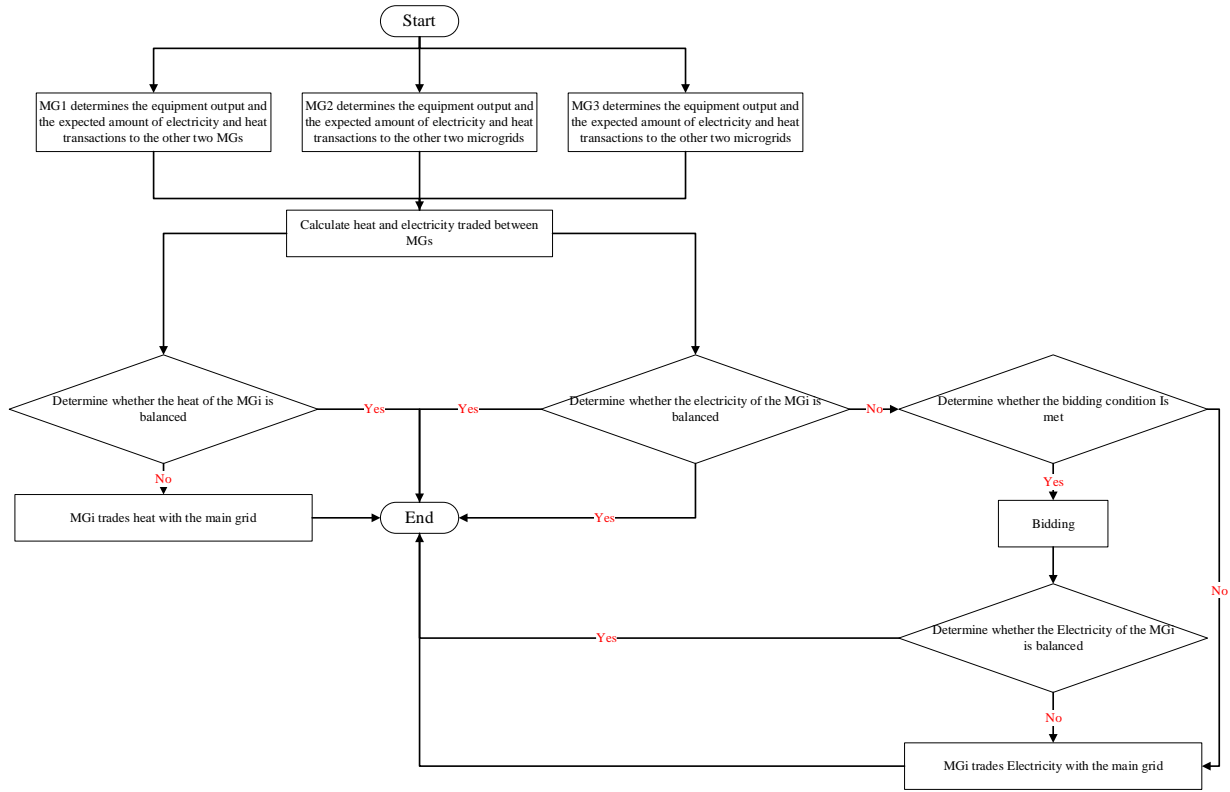

**Fig. A4** Flowchart of the paper

**References**

[1]. F Dong; L Shi. Regional differences study of renewable energy performance: A case of wind power in China[J]. Journal of Cleaner Production 2019;233:490-500

[2]. Merabet A, Al-Durra A, El-Saadany E F. Energy management system for optimal cost and storage utilization of renewable hybrid energy microgrid[J]. Energy Conversion and Management, 2022, 252: 115116.

[3]. Vand B, Ruusu R, Hasan A, et al. Optimal management of energy sharing in a community of buildings using a model predictive control[J]. Energy Conversion and Management, 2021, 239: 114178.

[4]. Bischi A, Taccari L, Martelli E, et al. A detailed MILP optimization model for combined cooling, heat and power system operation planning[J]. Energy 2014;74:12e26.

[5]. Rezaei N, Pezhmani Y. Optimal islanding operation of hydrogen integrated multi-microgrids considering uncertainty and unexpected outages[J]. Journal of Energy Storage 2022; 49: 104142.

[6]. dos Santos Neto P J, Barros T A S, Silveira J P C, et al. Power management techniques for grid-connected DC microgrids: A comparative evaluation[J]. Applied Energy; 2020;269: 115057.

[7]. Nelson J, Johnson N G, Fahy K, et al. Statistical development of microgrid resilience during islanding operation[J]. Applied Energy 2020; 279: 115724.

[8]. Jiang J, Zhou R, Xu H, et al. Optimal sizing, operation strategy and case study of a grid-connected solid oxide fuel cell microgrid[J]. Applied Energy 2022; 307: 118214.

[9]. Energy.Gov. About the Grid Modernization Initiative. Available online: https://www.energy.gov/grid-modernization-initiative

(accessed on 10 January 2019).

[10]. Zhao, B. Xue, M. Zhang, X. Wang, C. Zhao, J. An MAS based energy management system for a stand-alone microgrid at high altitude[J]. Applied Energy 2015; 143: 251–261.

[11]. Bacha, S. Picault, D. Burger, B. Etxeberria-Otadui, I. Martins, J. Photovoltaics in Microgrids: An overview of grid integration and energy management aspects[J]. IEEE Industrial Electronics Magazine 2015; 9: 33–46.

[12]. Valencia F, Collado J, Sáez D, et al. Robust energy management system for a microgrid based on a fuzzy prediction interval model[J]. IEEE Transactions on Smart Grid 2015; 7(3): 1486-1494.

[13]. Li Y, Han M, Shahidehpour M, et al. Data-driven distributionally robust scheduling of community integrated energy systems with uncertain renewable generations considering integrated demand response[J]. Applied Energy, 2023, 335: 120749.

[14]. Gupta R A, Gupta N K. A robust optimization based approach for microgrid operation in deregulated environment[J]. Energy Conversion and Management, 2015, 93: 121-131.

[15]. Mazidi M, Zakariazadeh A, Jadid S, et al. Integrated scheduling of renewable generation and demand response programs in a microgrid[J]. Energy Conversion and Management, 2014, 86: 1118-1127.

[16]. Abunima H, Park W H, Glick M B, et al. Two-stage stochastic optimization for operating a renewable-based microgrid[J]. Applied Energy 2022; 325: 119848.

[17]. Z. Wang，Q．Bian，H．Xin．A distributionally robust co-ordinated reserve scheduling model considering CVaR-based wind power reserve requirements[J]．IEEE Transactions on Sustainable Energy 2016; 7(2) :625-636．

[18]. Stoppato A, Cavazzini G, Ardizzon G, et al. A PSO (particle swarm optimization)-based model for the optimal management of a small PV (Photovoltaic)-pump hydro energy storage in a rural dry area[J]. Energy, 2014, 76: 168-174.

[19]. Vitale F, Rispoli N, Sorrentino M, et al. On the use of dynamic programming for optimal energy management of grid-connected reversible solid oxide cell-based renewable microgrids[J]. Energy 2021;2:120304.

[20]. Nawaz A, Zhou M, Wu J, et al. A comprehensive review on energy management, demand response, and coordination schemes utilization in multi-microgrids network[J]. Applied Energy, 2022, 323: 119596.

[21]. Xu D, Zhou B, Chan K W, et al. Distributed multienergy coordination of multimicrogrids with biogas-solar-wind renewables[J]. IEEE Transactions on Industrial Informatics, 2018, 15(6): 3254-3266.

[22]. Lan Y, Guan X, Wu J. Online decentralized and cooperative dispatch for multi-microgrids[J]. IEEE Transactions on Automation Science and Engineering, 2019, 17(1): 450-462.

[23]. Xu Y, Ye S, Qin Z, et al. A coordinated optimal scheduling model with Nash bargaining for shared energy storage and Multi-microgrids based on Two-layer ADMM[J]. Sustainable Energy Technologies and Assessments, 2023, 56: 102996.

[24]. Li J, Zhang C, Xu Z, et al. Distributed transactive energy trading framework in distribution networks[J]. IEEE Transactions on Power Systems, 2018, 33(6): 7215-7227.

[25]. Yutao Ju, Xi Chen, Jiawei Li, et al. Active and Reactive Power Coordinated Optimal Dispatch of Networked Microgrids Based on Distributed Deep Reinforcement Learning[J]. Automation of Electric Power Systems, 2023, 47(01):115-225.

[26]. Chen T, Bu S, Liu X, et al. Peer-to-peer energy trading and energy conversion in interconnected multi-energy microgrids using multi-agent deep reinforcement learning[J]. IEEE transactions on smart grid, 2021, 13(1): 715-727.

[27]. Wang Z, Zhang L, Tang W, et al. Equilibrium allocation strategy of multiple ESSs considering the economics and restoration capability in DNs[J]. Applied Energy 2022; 306: 118019.

[28]. Wu C, Gu W, Bo R, et al. Energy trading and generalized Nash equilibrium in combined heat and power market[J]. IEEE Transactions on Power Systems, 2020; 35(5): 3378-3387.

[29]. Hu M, Wang Y W, Xiao J W, et al. Multi-energy management with hierarchical distributed multi-scale strategy for pelagic islanded microgrid clusters[J]. Energy, 2019, 185: 910-921.

[30]. Nash J. Non-cooperative games[J]. Annals of mathematics, 1951: 286-295.

[31]. Liu H. Multi-agent Cooperative Control Based on Potential Game Theory[D]. Liaoning: University of Science and Technology Liaoning.

[32]. Ma Z, Zhou X, Shang Y, et al. Exploring the concept key technologies and development model of energy internet[J]. Power

System Technology, 2015; 39(11):3014-3022.

[33]. Shu Y, Zhang Z, Guo J, et al. Study on key factors and solution of renewable energy accommodation[J]. Proceedings of the CSEE 2017;37(1):1-8.

[34]. Ma Z, Zhou X, Shang Y, et al. Form and development trend of future distribution system[J]. Proceedings of the CSEE 2015; 35(6): 1289-1298.

[35]. Bui V H, Hussain A, Kim H M. Double deep Q-learning-based distributed operation of battery energy storage system considering uncertainties[J]. IEEE Transactions on Smart Grid 2019; 11(1): 457-469.

[36]. Konda V, Tsitsiklis J. Actor-critic algorithms[J]. Advances in neural information processing systems, 1999, 12.

[37]. Lillicrap TP, Hunt J J, Pritzel A, et al. Continuous control with deep reinforcement learning[J]. arXiv preprint arXiv:1509.02971, 2015.

[38]. Zhang F, Yang Q, An D. CDDPG: A deep-reinforcement-learning-based approach for electric vehicle charging control[J]. IEEE Internet of Things Journal, 2020, 8(5): 3075-3087.

[39]. Lei L, Tan Y, Dahlenburg G, et al. Dynamic energy dispatch based on deep reinforcement learning in IoT-driven smart isolated microgrids[J]. IEEE internet of things journal 2020;8(10): 7938-7953.

[40]. Schulman J, Wolski F, Dhariwal P, et al. Proximal policy optimization algorithms[J]. arXiv preprint arXiv:1707.06347, 2017.

[41]. Guo C, Wang X, Zheng Y, et al. Real-time optimal energy management of microgrid with uncertainties based on deep reinforcement learning[J]. Energy, 2022, 238: 121873.

[42]. Yi Z, Xu Y, Wang X, et al. An improved two-stage deep reinforcement learning approach for regulation service disaggregation in a virtual power plant[J]. IEEE Transactions on Smart Grid, 2022.

[43]. Biemann M, Scheller F, Liu X, et al. Experimental evaluation of model-free reinforcement learning algorithms for continuous HVAC control[J]. Applied Energy, 2021, 298: 117164.

[44]. Haarnoja T, Zhou A, Abbeel P, et al. Soft actor-critic: Off-policy maximum entropy deep reinforcement learning with a stochastic actor[C]//International conference on machine learning. PMLR, 2018: 1861-1870.

[45]. Lowe R, Wu Y I, Tamar A, et al. Multi-agent actor-critic for mixed cooperative-competitive environment[J]. Advances in neural information processing systems, 2017, 30.

[46]. Fang X, Zhao Q, Wang J, et al. Multi-agent deep reinforcement learning for distributed energy management and strategy optimization of microgrid market[J]. Sustainable Cities and Society, 2021, 74: 103163.

[47]. Xi L, Sun M, Zhou H, et al. Multi-agent deep reinforcement learning strategy for distributed energy[J]. Measurement, 2021, 185: 109955.

[48]. A. Yona, T. Senjyu and T. Funabashi. Application of recurrent neural network to short-term-ahead generating power forecasting for photovoltaic system[C]. 2007 IEEE Power Engineering Society General Meeting, 2007, pp. 1-6, doi: 10.1109/PES.2007.386072.

[49]. Cai Z. Corrected Numerical Weather Prediction-BP Neural Network Basd Wind Power Short-term Predictions Research[D]. Zhejiang: Zhejiang University.

[50]. Xu Q，Zeng A，Wang K，et al．Day-ahead optimized economic dispatching for combined cooling，heating and power in micro energy-grid based on hessian interior point method[J]. Power System Technology, 2016, 40(6):1657-1665.

[51]. S. Deng. Optimal Configuration and Operation of an Energy Hub Considering Off-design Characteristics of Generation Units[D]. Guangzhou：South China University of Technology.

[52]. Liu S, Qian Y, Li D, et al. Multi-Scenario scheduling optimisation for a novel Double-Stage ammonia absorption refrigeration system incorporating an organic Rankine cycle[J]. Energy Conversion and Management, 2022, 270: 116170.

[53]. Wei T. Design and Operation Characteristics of Solar Energy Utilization System in Alpine Region[D]．Baoding：North China Electric Power University.

[54]. Lai K. Investigation of a biomass-CCHP(combined cooling, heating and power) system based on the downdraft fixed bed gasifier characteristics under all conditions[D]. Shanghai:Shanghai Jiao Tong University.

[55]. Ju S. Research on Operation Optimization of Concentrated Solar Power Thermal Power Generation System[D]. Jilin: Northeast Electric Power University.

[56]. Meybodi M A, Behnia M. A study on the optimum arrangement of prime movers in small scale microturbine-based CHP systems[J]. Applied Thermal Engineering, 2012, 48: 122-135.

[57].CHEN Wanqing, MU Yunfei, JIA Hongjie, et al. Operation Optimization Method for Regional Integrated Energy System Considering Part-load Performances of Devices[J]. Power System Technology, 2021, 45(3): 951-958.

[58].Guan T, Zuo P, Sun S, et al. Degradation mechanism of LiCoO2/mesocarbon microbeads battery based on accelerated aging tests[J]. Journal of Power Sources, 2014, 268: 816-823.

[59]. Haarnoja T, Zhou A, Hartikainen K, et al. Soft actor-critic algorithms and applications[J]. arXiv preprint arXiv:1812.05905, 2018.